# Watch Your Adjoints! Lack of Mesh Convergence in Inviscid Adjoint Solutions

Carlos Lozano*

*National Institute for Aerospace Technology (INTA), Torrejón de Ardoz, Spain, 28850*

It has been long known that 2D and 3D inviscid adjoint solutions are generically singular at sharp trailing edges. In this paper, a concurrent effect is described by which wall boundary values of 2D and 3D inviscid continuous and discrete adjoint solutions based on lift and drag are strongly mesh dependent and do not converge as the mesh is refined. Various numerical tests are performed to characterize the problem. Lift-based adjoint solutions are found to be affected for any flow condition, while drag-based adjoint solutions are affected for transonic lifting flows. A (laminar) viscous case is examined as well, but no comparable behavior is found, which suggests that the issue is exclusive to inviscid flows. It is argued that this behavior is caused by the trailing edge adjoint singularity.

## Nomenclature

### Latin letters

| | | |
|---|---|---|
| $\vec{F}_U$ | = | inviscid Jacobian vector |
| $\vec{d}$ | = | force direction vector |
| $E$ | = | total energy |
| $\vec{F}$ | = | Euler Cartesian flux vector |
| $H$ | = | total enthalpy |
| $I$ | = | objective function |
| $L$ | = | Lift |
| $M_\infty$ | = | free-stream Mach number |

* Research Scientist, Computational Aerodynamics Group, Carretera de Ajalvir, km. 4, Torrejón de Ardoz (Spain).

$\vec{n}$ = unit normal vector

$p$ = pressure

$Re$ = Real part of complex number

$S$ = wall boundary

$S^{\infty}$ = “far field”

$U$ = conservative flow variables

$\vec{V}$ = velocity vector

$\hat{x}, \hat{y}$ = Cartesian unit vectors

$x,y$ = Cartesian coordinates

$z$ = complex coordinate

**Greek letters**

$\alpha$ = angle of attack

$\delta$ = first variation

$\zeta$ = complex coordinate

$\theta$ = polar angle

$\rho$ = density

$\vec{\varphi}$ = adjoint velocity vector

$\psi$ = adjoint variables

$\Lambda$ = adjoint stream function

$\Psi$ = stream function

$\Omega$ = fluid domain

**Mathematical symbols**

$\nabla$ = nabla (gradient) operator

$\partial$ = partial derivative

**Superscripts**

$T$ = transpose of a matrix

$\rightarrow$ = vector quantity

$\overline{\phantom{x}}$ = complex conjugation

**Subscripts**

$x,y$ = Cartesian components of a vector

$r,\theta$ = Derivative with respect to radial/angular polar coordinates

## I. Introduction

THIRTY years ago, Jameson sparked a revolution by bringing adjoint methods into the arena of aerodynamic design optimization [1, 2]. Since then, the adjoint approach has been extended to cover a wide variety of CFD-based applications including shape design, flow control, uncertainty quantification, error estimation and mesh adaptation. The solution of the adjoint equations links the sensitivity of a given functional or cost function to perturbations of the flow such as, for example, shape deformations in design applications or discretization errors in error estimation applications. In the former case, the adjoint solution can be used to compute the gradients of the cost function with respect to the design variables at a cost that is independent of the size of the design space.

Starting with Jameson's groundbreaking work, much effort has been put in developing adjoint-based optimization methods [3, 4, 5, 6, 7, 8, 9, 10, 11], while relatively little attention has been paid to the adjoint solutions themselves (see however [12, 13, 14, 15, 16, 17, 18] ). It has been long known that 2D and 3D inviscid adjoint solutions are singular along stagnation streamlines [12] and sharp trailing edges, both analytically [19] [20] and numerically (see, for example, Fig. 4 in [12]). These singularities are clearly noticeable in adjoint-based mesh adaptation applications [21], as nodes in adjoint-adapted meshes tend to significantly cluster around those areas. In the case of the trailing edge singularity, extreme adaptive refinement is required to recover optimal convergence rates [22]. In design applications, on the other hand, the trailing edge adjoint singularity significantly degrades the accuracy of reduced adjoint gradient formulations such as [5] where only surface perturbations are considered. For example, it has been noted that the singularity prevents the applicability of boundary formulas based on shape calculus [23] that only consider the normal part of the surface deformation. Most importantly, it has been shown [24, 4] that due to the singularity, volume grid sensitivities cannot be ignored and do play a crucial role in the computation of gradients.

In this paper we describe a related problem first observed in [17] and [25] when performing a mesh convergence study of an apparently innocent (drag) adjoint solution corresponding to transonic, inviscid, steady flow past a NACA0012 airfoil with $M_\infty = 0.8$ and $\alpha = 1.25^\circ$ . The problem is clearly spotted in Fig. 3 as a notorious mesh divergence of the adjoint values on the airfoil surface. This lack of mesh convergence hampers the interpretation of numerical results, making it difficult to compare results obtained with different codes and/or meshes, and can become an issue for mesh adaptation, as the growing size of wall adjoint variables may result in excessive refinement towards the wall (see for example Fig. 7 in [22]). We will show that the problem is not limited to two-dimensional transonic cases or to the drag functional but it is rather a generic problem for inviscid solutions past bodies with sharp trailing edges, which is likely caused by the adjoint singularity at the trailing edge. Likewise, the problem appears in solutions computed with both continuous and discrete adjoint solvers and is, thus, different from other numerical artifacts reported in the literature that plague adjoint solutions (usually from discrete adjoint approaches) due to the lack of dual consistency in the numerical scheme [24, 26, 27, 28, 29, 30, 31]. This distinction is important, as it directly affects what one should expect to see upon refining the mesh. If the adjoint discretization is a consistent approximation to the dual problem (and continuous adjoint discretizations usually are), the method is expected to converge towards the solution of the adjoint PDE with mesh refinement except, of course, at singularities of the adjoint equations. Here the quasi-1d adjoint solution offers some insight: the adjoint solution does have logarithmic and jump singularities [13, 17] at choked throats, depending on the cost function, but at least for consistent adjoint discretizations the numerical solution converges to the analytic solution (that can be computed exactly using Giles and Pierce's Green function approach [13]) as the mesh is refined.

A priori, a similar behavior should be expected in 2D: point-wise convergence to the "analytic" adjoint solution, whatever it might be, for consistent discretizations except at singularities. From the work of Giles and Pierce [12] we know that the adjoint solution is expected to be smooth (and singularity-free) at generic points of walls. Likewise, in transonic flows the adjoint solution is regular at generic points along shocks and sonic lines, although numerical adjoint solutions for flows with primal shocks may be non-unique [32, 33, 34] unless an internal shock boundary condition [13] is enforced or sufficient numerical smoothing is applied. On the other hand, adjoint solutions are singular at sharp trailing edges and along the incoming stagnation streamline. Likewise, transonic adjoint solutions also typically exhibit a discontinuity along the supersonic characteristic emanating from the root of the shock that reflects off the sonic line creating a triangular structure [22, 35] which is clearly visible in adjoint-adapted meshes. It has also been noted that adjoint solutions are also discontinuous at contact lines [36], so the slip lines/surfaces that are present in inviscid transonic rotational primal flow past airfoils and wings with sharp trailing edges are also candidates for an adjoint discontinuity. Finally, a singularity is also apparent in transonic solutions at the root of the sonic line (see e.g. Fig. 4 of [12]). At these locations, clean mesh convergence of the adjoint solution should not be expected.

We will devote the remainder of the paper to characterize the problem and try to figure out its cause.

## II.Statement of the Problem

### A. The Adjoint Equations

We begin by recalling a few facts regarding the inviscid adjoint equations. We will focus, for definiteness, on steady, two-dimensional, inviscid flow on a domain $\Omega$ with far-field boundary $S^{\infty}$ and wall boundary *S* (typically an airfoil profile). The flow is governed by the Euler equations $\nabla \cdot \vec{F} = 0$, where $\vec{F} = (\rho\vec{v}, \rho\vec{v}v_x + p\hat{x}, \rho\vec{v}v_y + p\hat{y}, \rho\vec{v}H)^T$ is the flux vector and $\rho, \vec{v}, p, E, H$ the fluid's density, velocity, pressure, total energy and enthalpy, respectively. The adjoint equations are defined with respect to a functional of the flow variables, or cost function, that we take to be the force exerted by the fluid on the boundary *S* measured along a direction $\vec{d}$

$$I(S) = \int_S p(\vec{n}_S \cdot \vec{d})dS \tag{1}$$

(where $\vec{n}_S$ is the outward-pointing normal vector to the solid surface). With appropriate choices of $\vec{d}$, eq. (1) can represent the drag or lift on the airfoil.

The corresponding adjoint state $\psi = (\psi_1, \psi_2, \psi_3, \psi_4)^T$ obeys the (adjoint) equation

$$\vec{F}_U^{\,T} \cdot \nabla\psi = 0 \qquad \text{in } \Omega \tag{2}$$

with the following wall and far-field boundary conditions

$$\begin{aligned} \vec{\varphi} \cdot \vec{n}_S = \vec{n}_S \cdot \vec{d} \qquad & \text{on } S \\ \psi^T (\vec{F}_U \cdot \vec{n}_{S^{\infty}})\delta U = 0 \qquad & \text{on } S^{\infty} \end{aligned} \tag{3}$$

where $\vec{\varphi} = (\psi_2, \psi_3)$ is the adjoint velocity vector. (If the flow solution contains a shock, the above equations need to be modified to account for the appropriate adjoint shock conditions [15, 17]). The adjoint equations (2)-(3) can be discretized for numerical computation (continuous adjoint approach); alternatively, the adjoint system can be derived directly from the discretized flow equations (discrete adjoint approach).

### B. The singularity at the trailing edge

In [19, 20], Giles and Pierce constructed via conformal mapping a 2D potential adjoint solution that exhibited a singularity at the sharp (cusped) trailing edge of an airfoil (Fig. 1). Their solution corresponds to non-lifting potential flow past a symmetric Joukowski airfoil obtained by applying the Joukowski transformation $z = \zeta + \zeta^{-1}$ to the region bounded by two concentric circumferences $C_1$ and $C_2$ of radii $R_1 = 1.1$ and $R_2 = 3$ centered at $(-0.1, 0)$ on the $\zeta$-plane. In polar coordinates $\zeta = -0.1 + r\,e^{i\theta}$, the stream function is

$$\Psi(r,\theta) = \frac{r^2 - R_1^2}{r}\sin\theta$$

The adjoint problem is defined with respect to the cost function

$$-\int_0^{2\pi} \sin\theta\, \Psi_r\big|_{r=R_1}\, d\theta$$

(where $\Psi_r = \partial\Psi / \partial r$). The corresponding adjoint stream function $\Lambda$ obeys $\nabla^2\Lambda = 0$ throughout the domain, with boundary conditions $\Lambda\big|_{r=R_1} = \sin\theta / R_1$ and $\Lambda\big|_{r=R_2} = 0$, and can be found to be

$$\Lambda(r,\theta) = -\frac{1}{R_2^{\,2} - R_1^{\,2}}\left(r - \frac{R_2^{\,2}}{r}\right)\sin\theta \qquad (4)$$

In the *z*-plane, the adjoint velocities

$$\psi_x - i\psi_y = -\frac{1}{R_2^{\,2} - R_1^{\,2}}\left[1 - \frac{R_2^{\,2}}{(\zeta + 0.1)^2}\right]\left(1 - \frac{1}{\zeta^2}\right)^{-1}$$

are singular at the trailing edge $\zeta_{t.e.} = 1$. The singularity has the same origin as the corresponding one in the primal flow that is avoided with the Kutta condition.

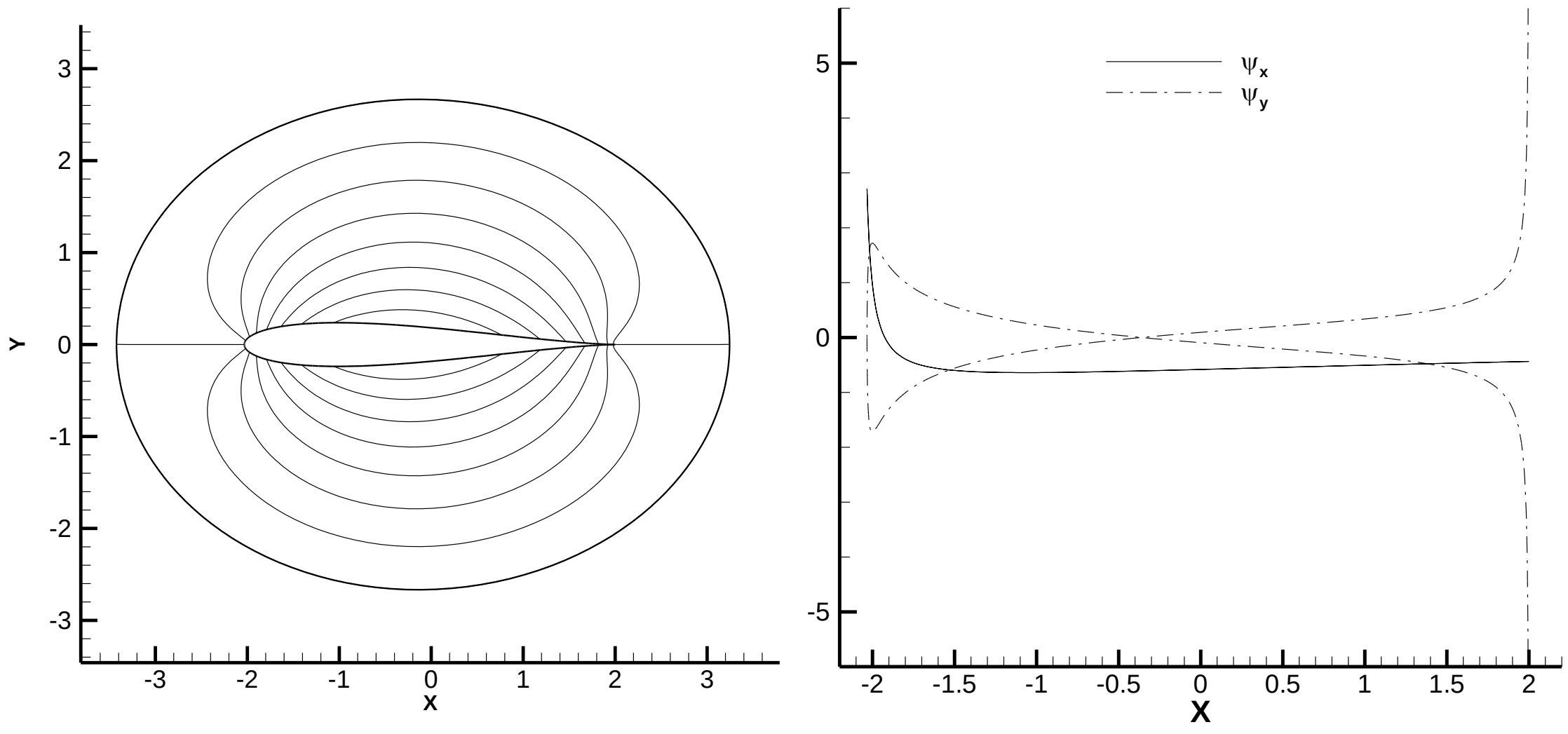

(a) (b)

**Fig. 1 (a) Contours of adjoint stream function and (b) adjoint velocity components on the airfoil profile for the potential solution of [19, 20].**

We can generalize the above solution to potential flow with circulation and a different cost function such as lift. The stream function is

$$\Psi(r,\theta) = \left(r - \frac{R_1^2}{r}\right)\sin(\theta-\alpha) + 2R_1 \sin\alpha \log\frac{r}{R_1}$$

(where $\alpha$ is the angle of attack), in terms of which the lift can be represented as

$$L = R_1 \oint_{C_1} \Psi_r \, d\theta$$

The adjoint stream function obeys $\nabla^2\Lambda = 0$ with boundary conditions $\Lambda|_{r=R_1} = -1$ and $\Lambda|_{r=R_2} = 0$, which yields

$$\Lambda(r,\theta) = \frac{\log\frac{r}{R_2}}{\log\frac{R_2}{R_1}} \tag{5}$$

On the physical plane, the complex adjoint velocities

$$\psi_x - i\psi_y = i\frac{1}{\log\frac{R_2}{R_1}}\left(1-\frac{1}{\zeta^2}\right)^{-1}\frac{1}{\zeta+0.1}$$

are again singular at the trailing edge, as shown in Fig. 2.

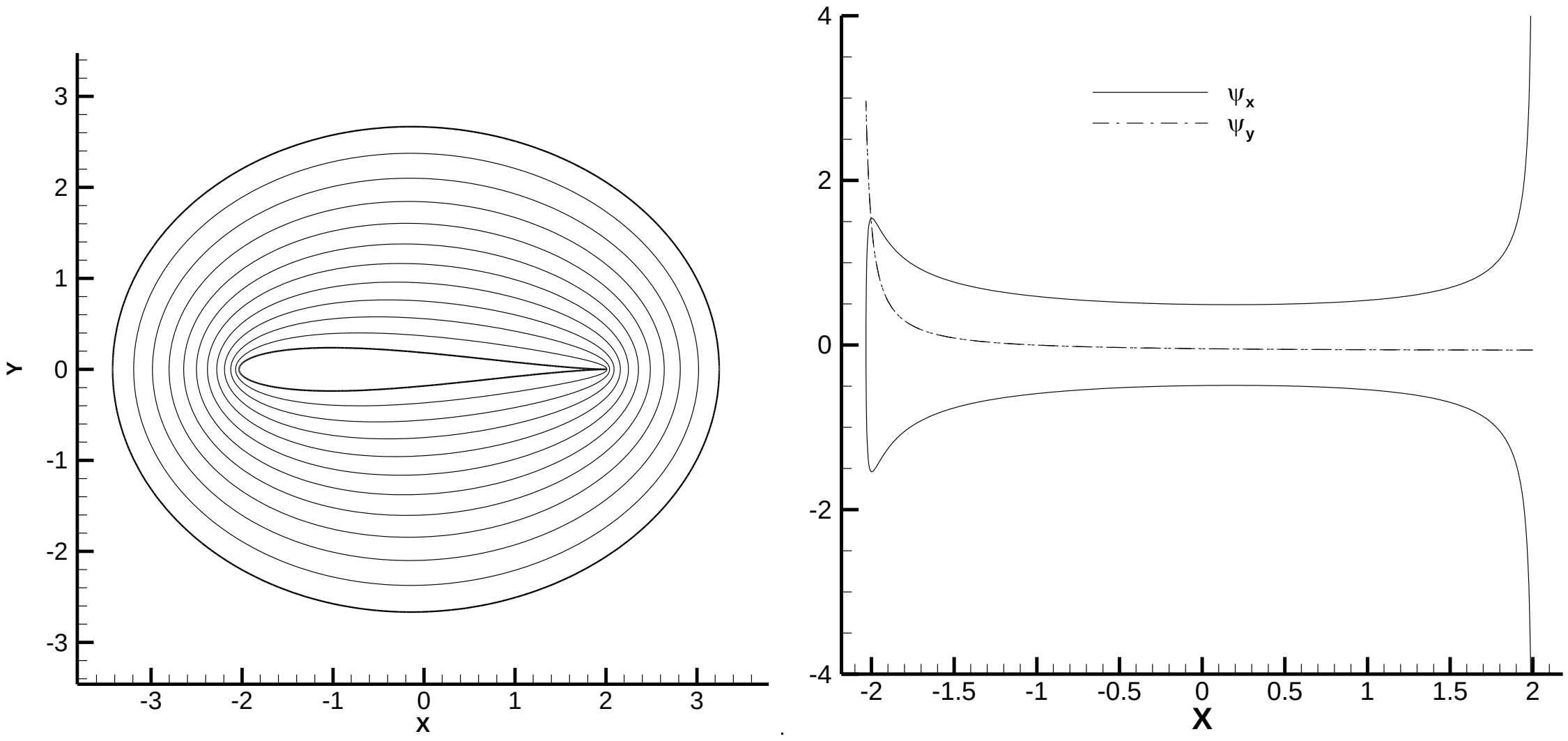

**(a)** **(b)**

**Fig. 2 (a) Contours of the lift adjoint stream function and (b) lift adjoint velocity components on the airfoil profile for lifting potential flow with $\alpha = 2^\circ$.**

There are several points to note here. First, the singularity is associated to the adjoint "flow" turning around the sharp trailing edge, as in (primal) potential flow without a Kutta condition. Second, and related to the previous one, no "adjoint" Kutta condition is possible here, as neither (5) (nor (4)) admits extra terms to force an adjoint stagnation point at the trailing edge on the $\zeta$-plane. Finally, the solution is perfectly smooth along the remainder of the wall. This matches Giles and Pierce's observation in [12] and makes it clear that there is no reason to expect any anomalous behavior of adjoint solutions at generic points of solid walls.

For inviscid transonic flow, no analytic solution is available but, as already emphasized in the introduction, the singularity appears to be also present, at least in numerical computations [12, 21]. Fig. 3 plots the drag density adjoint variable on the surface of a NACA0012 airfoil for transonic, inviscid flow with $M_\infty = 0.8$ and $\alpha = 1.25^\circ$ computed with DLR's discrete adjoint Tau solver [37] on a set of 6 sequentially refined triangular meshes with sizes ranging from ~ 3000 to ~3 million nodes, respectively. The farfield has a radius of 50 chord lengths in all cases, and each mesh is obtained by uniform refinement of the previous one (that is to say, by splitting in half every mesh edge). When edges are split on the airfoil's surface, the position of the new points is adjusted using cubic splines interpolation such that the new discrete surface follows the original geometry.

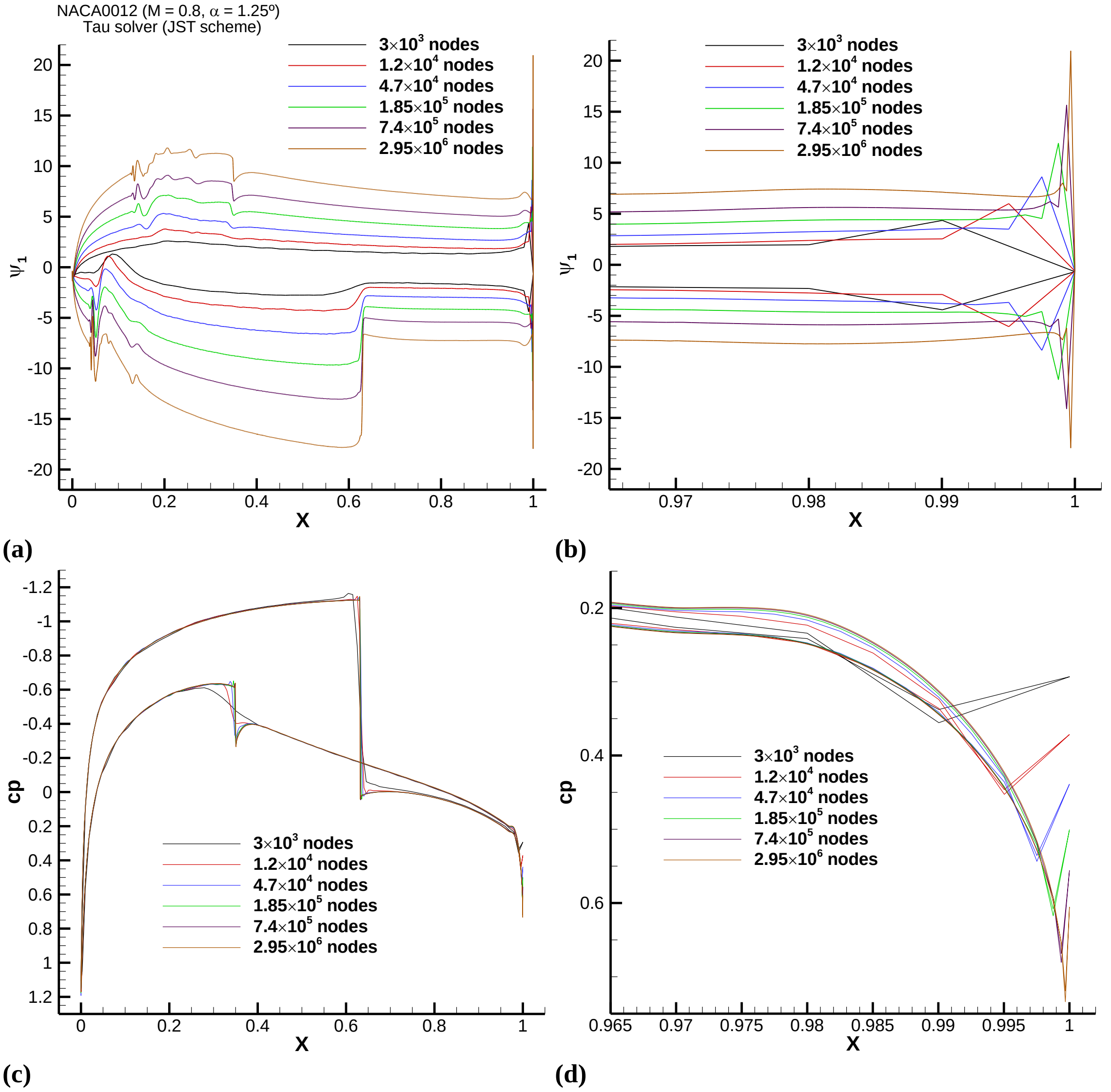

**Fig. 3 NACA0012 airfoil with $M_\infty = 0.8$ and $\alpha = 1.25^o$. (a) and (b): Drag adjoint density solution on the airfoil surface on 6 meshes with the Tau solver. (c) and (d): Surface pressure coefficients.**

At the specified conditions, the flow solution has a shock at $x/c \approx 0.63$ on the suction side of the airfoil (with the maximum local Mach number reaching approximately 1.4 on the supersonic side of the shock), and a weaker one on the pressure side. In this case, the actual value of the numerical adjoint solution at the trailing edge is close to zero, but the singularity manifests itself in the large values of the adjoint variable near the trailing edge, which grow continually when the mesh around the trailing edge is refined (Fig. 3 (b) and also Fig. 4). This mesh dependence should be expected if there is a singularity at the trailing edge: when the mesh is refined, nodes get closer to the singularity and larger and larger values are obtained.

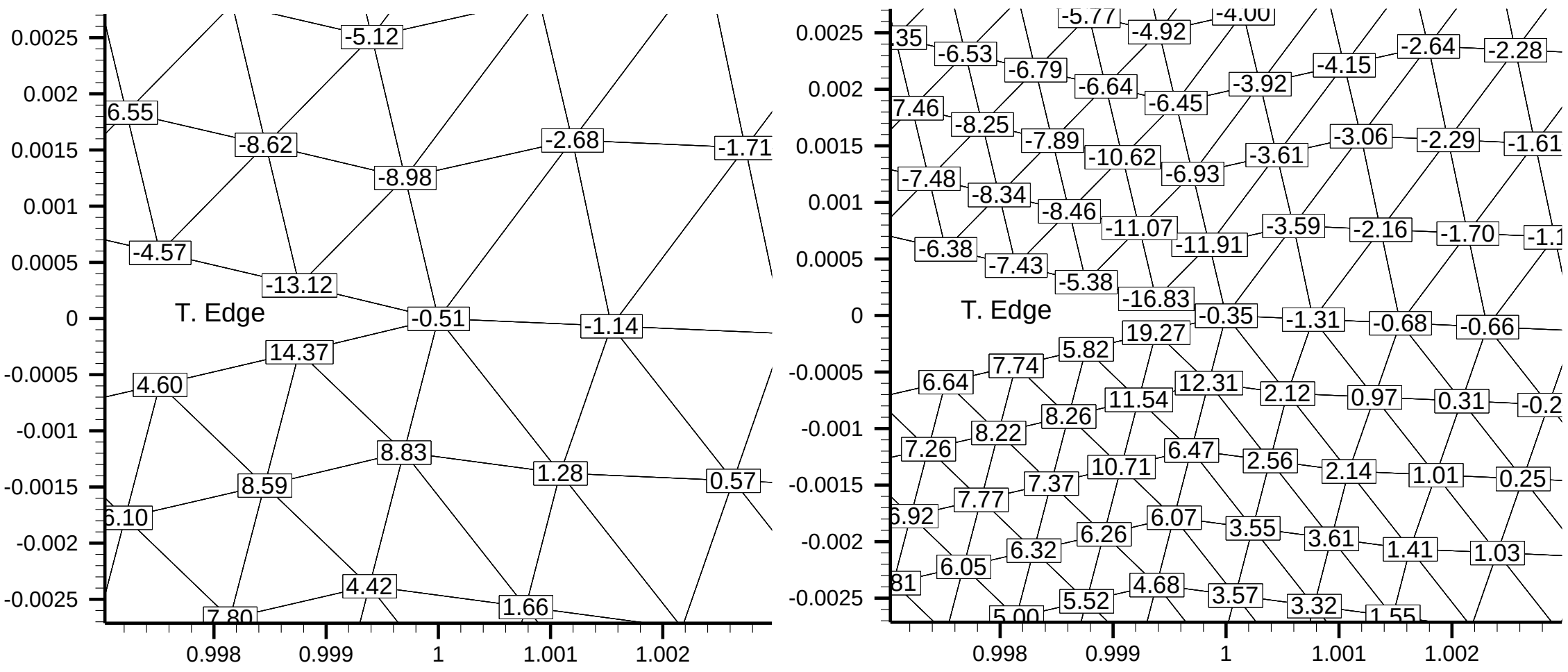

**(a)** **(b)**

**Fig. 4 NACA0012: $M_\infty = 0.8$ and $\alpha = 1.25^o$. Magnitude of the drag adjoint density for two sequentially refined meshes with 185K (a) and 740K nodes (c). (Blow-up near trailing edge)**

Although not shown, the remaining adjoint variables behave in a similar fashion. Likewise, changing the flow conditions or the cost function affects the singularity: for the NACA0012 airfoil, drag adjoint variables are not singular for subsonic (shock-free) flow or zero angle of attack, while lift adjoint variables are singular at any angle of attack and Mach number, as will be shown in section III.

All of the above is well-known, and simply reflects the sensitivity of the Kutta condition to changes in the geometry of the trailing edge [19]. What is not so well known, and is apparent in Fig. 3, is the very clear mesh divergence of the adjoint solution across the whole airfoil profile, with values growing continually as the mesh is refined. However, adjoint-based sensitivity derivatives (Fig. 5 ) are quite accurate [25] and do not reflect a comparable level of mesh dependence, which confirms the idea that validating adjoint solutions via the sensitivities may be misleading [38]. The adjoint wall boundary condition (3) is also fairly well behaved upon mesh refinement (Fig. 6).

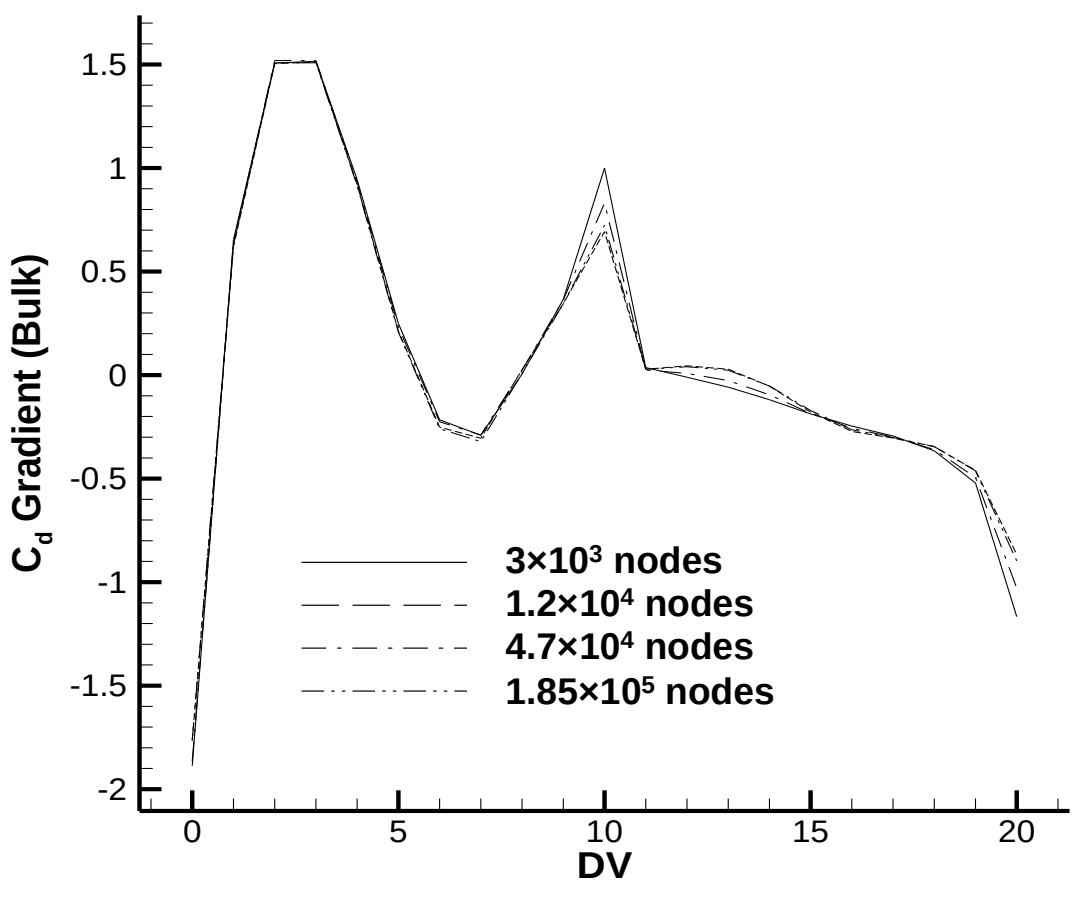

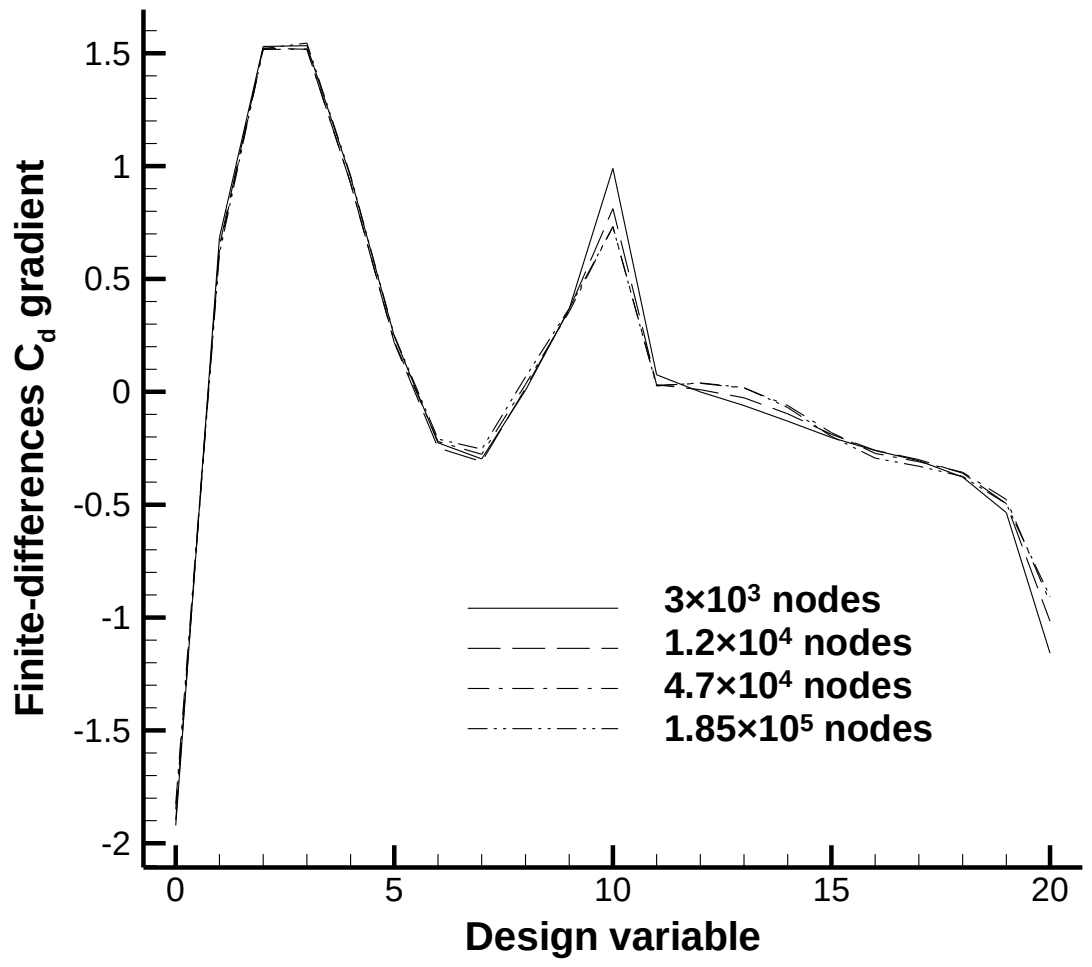

**(a)** **(b)**

**Fig. 5 NACA0012 airfoil with $M_\infty = 0.8$ and $\alpha = 1.25^\circ$. (a) Discrete adjoint drag gradients (b) Finite-differences gradients[†]**

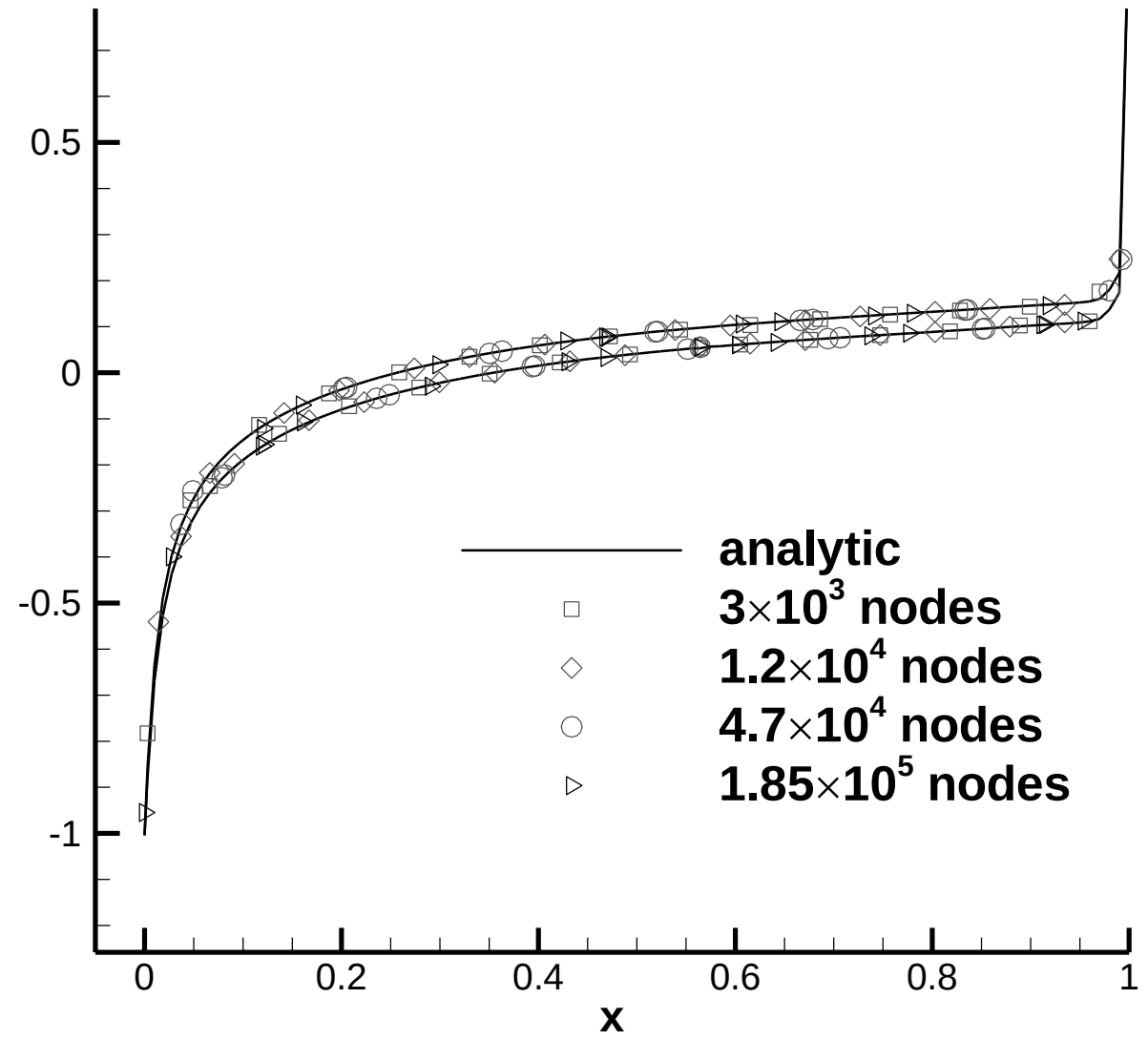

**Fig. 6 Numerical vs. analytic wall adjoint boundary condition for the NACA0012 airfoil with $M_\infty = 0.8$ and $\alpha = 1.25^\circ$**

[†] Design variables 1-10 correspond to local deformations (via Hicks-Henne functions) of the suction side of the airfoil from the leading edge (#1) to the trailing edge (#10). Variables 11-20 correspond to local deformations of the pressure side from leading (#11) to trailing (#20) edges. # 0 displaces the trailing edge along the vertical direction.

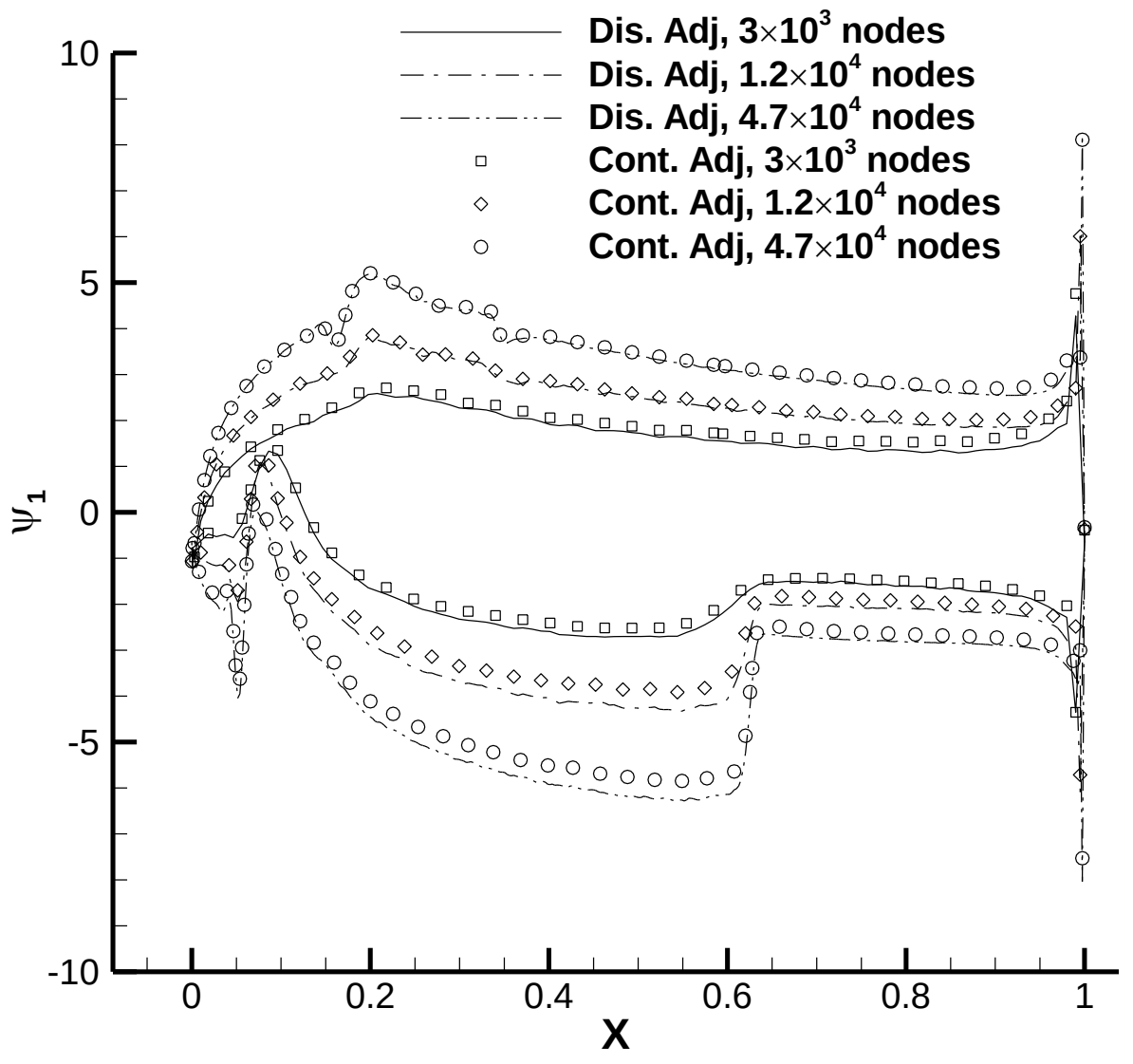

**Fig. 7 NACA0012 airfoil with $M_\infty = 0.8$ and $\alpha = 1.25^o$. Drag adjoint density solution on the airfoil profile on meshes 0-2 with Tau's continuous and discrete adjoint solvers.**

Fig. 3 also shows the $C_p$ distribution on the airfoil profile. While $C_p$ values at the trailing edge do change as the mesh is refined, we do not believe that this is the cause of the problem. The same behavior is observed in Fig. 8 which however corresponds to a case where adjoint variables do not show either a trailing edge adjoint singularity or a generalized mesh dependence issue.

The observed behavior is not exclusive to the chosen code or numerical settings. Similar trends are observed with Tau's continuous adjoint solver (Fig. 7), on different sets of meshes, with different schemes (a second-order central discretization with JST scalar dissipation [39] has been used for Fig. 3, although results with Roe's upwind solver [40] show similar trends). Likewise, the same behavior has been observed with Stanford University's SU2 finite-volume, unstructured continuous adjoint solver [41]. Likewise, ref. [42] reports an offset in surface drag adjoint values with respect to Tau computations for the transonic NACA case computed with the high-order spectral/hp element solver Nektar++.[3]

In section III it will be shown that the issue is strongly related to the trailing edge singularity, in such a way that both are simultaneously present or absent.

[3] The computed values also show a continuous variation with mesh refinement or order increase (D. Ekelschot, private communication).

# III. Numerical experiments

We have seen in Fig. 3 that for the transonic NACA0012 case at $M_\infty = 0.8$ and $\alpha = 1.25^\circ$, there is a lack of mesh convergence properties of the adjoint variables across the whole airfoil profile. Barring bugs or implementation issues, this behavior must have a numerical origin (as argued above, it certainly cannot be explained from the viewpoint of the adjoint p.d.e.) and it is important to know if it is related to the trailing edge singularity and whether it is specific to the flow conditions and cost function chosen. In order to characterize the problem, several test cases have been analyzed changing the flow conditions, the trailing edge geometry and the cost function. A 3D case will also be investigated, as well as a 2D viscous (laminar) case.

## A. 2D inviscid flows

*Effect of angle of attack*

Fig. 3 corresponds to a non-symmetric (lifting) transonic case, for which the Kutta condition results in a slip line (contact discontinuity) that emanates from the sharp trailing edge [43] and that is absent in the symmetric case. Reducing the angle of attack does seem to reduce (and even eliminate) both the singularity and the continuous mesh variation. This can be seen in Fig. 8 , where the surface value of the drag adjoint density for the NACA0012 airfoil with $M = 0.8$ and $\alpha = 0^\circ$ on 6 sequentially finer meshes is presented, clearly showing a totally different behavior at the trailing edge and across the airfoil. The solution depicted in Fig. 8 does still show a mesh divergence at around $x/c \approx 0.08$ that actually corresponds to the root of the sonic line of the primal flow, which appears to be a singularity of the adjoint solution that can also be spotted in Fig. 3, Fig. 17 and Fig. 22, for example.

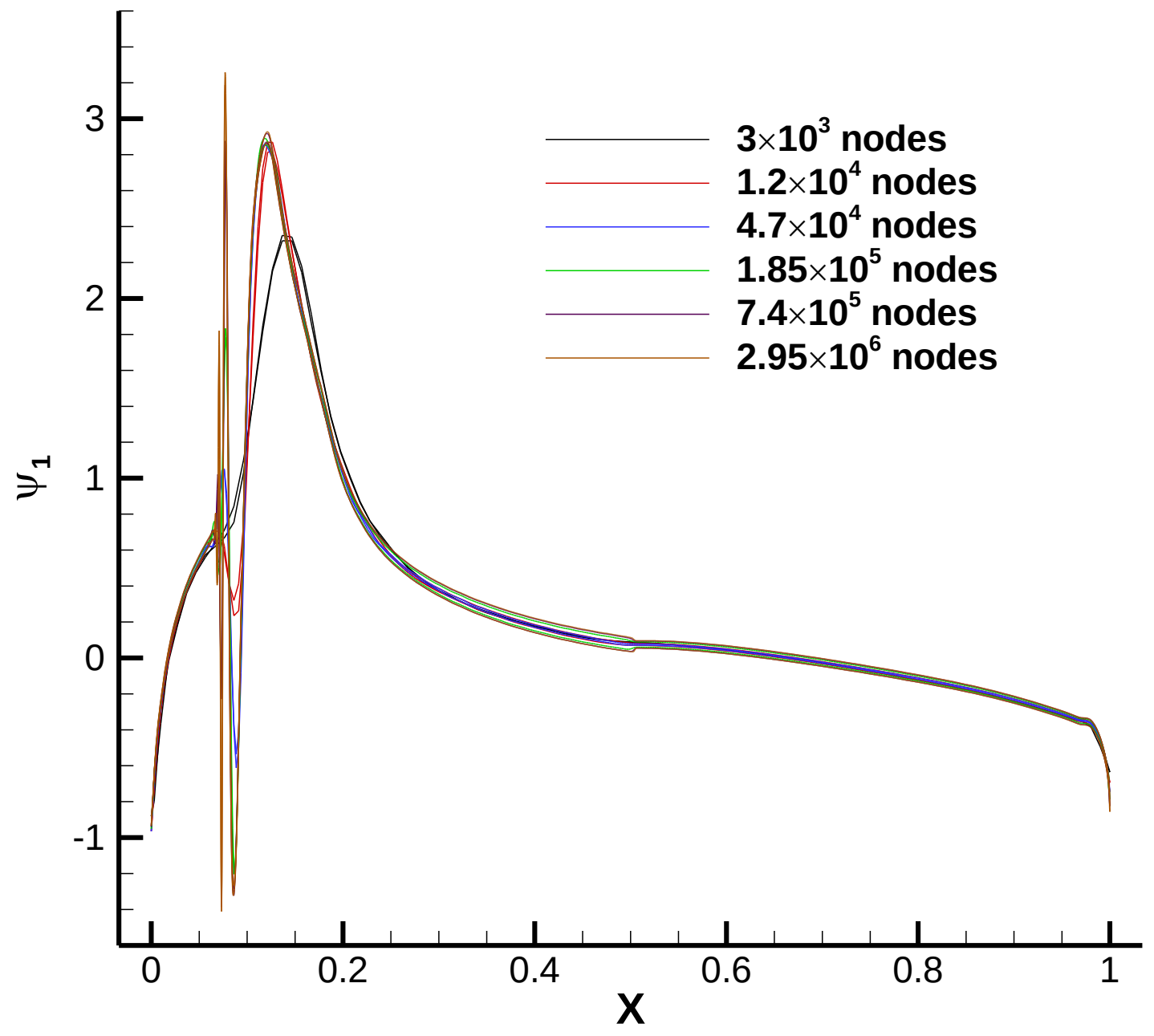

(a)

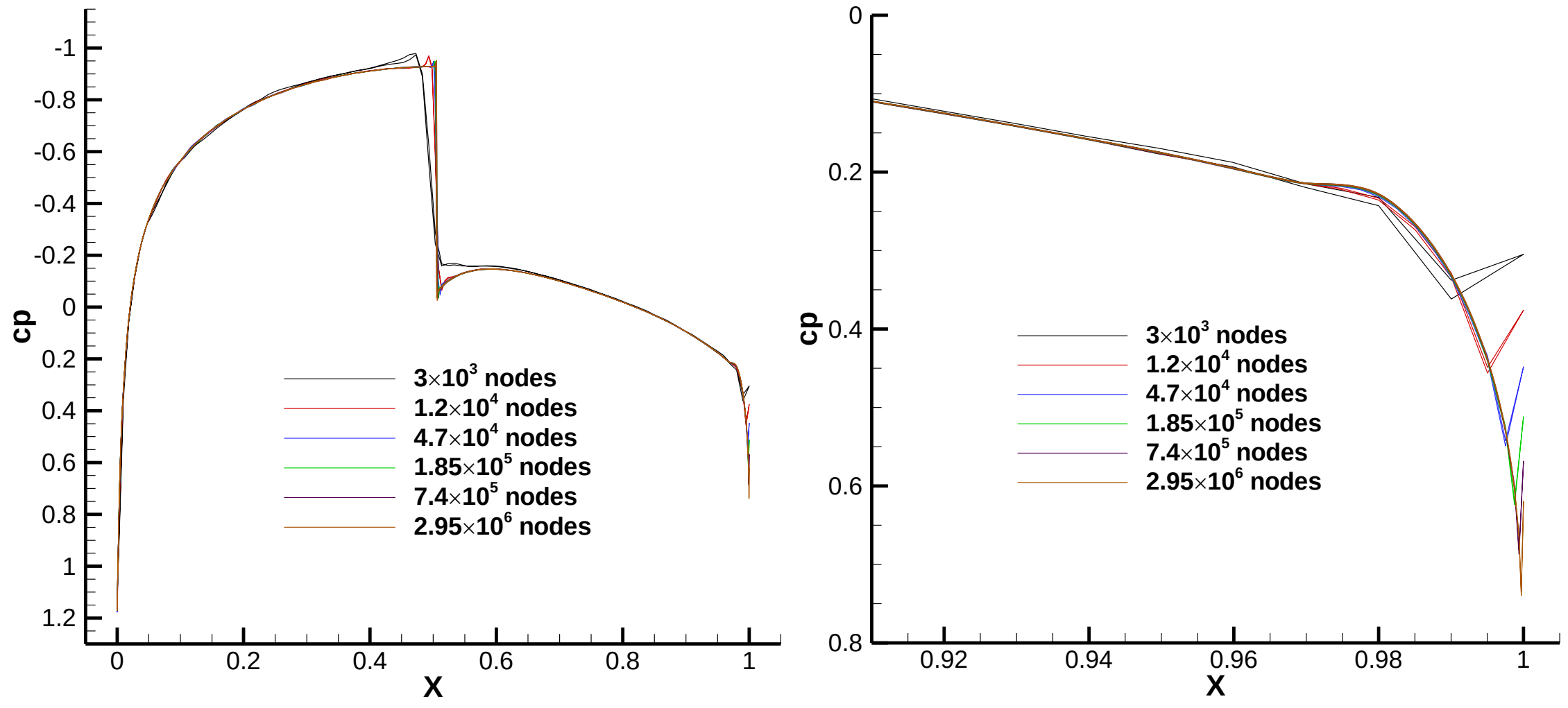

(b) (c)

**Fig. 8 NACA0012 airfoil with $M_\infty = 0.8$ and $\alpha = 0$ (Tau solver). (a) Drag adjoint solution. (b) Overview of pressure coefficient. (c) Zoom of the pressure coefficient near the trailing edge.**

Notice also that while $C_p$ values show a level of mesh dependence at the trailing edge comparable to that in Fig. 3, the behavior of the adjoint solution is completely different.

Further intuition can be gained by comparing the patterns of the adjoint momentum vector $\vec{\varphi} = (\psi_2, \psi_3)$ for both transonic NACA0012 cases. This is done in Fig. 9 , where it is shown that the singularity at the trailing edge is associated with the adjoint momentum "flow" turning around the trailing edge, as in the potential flow case analyzed above.

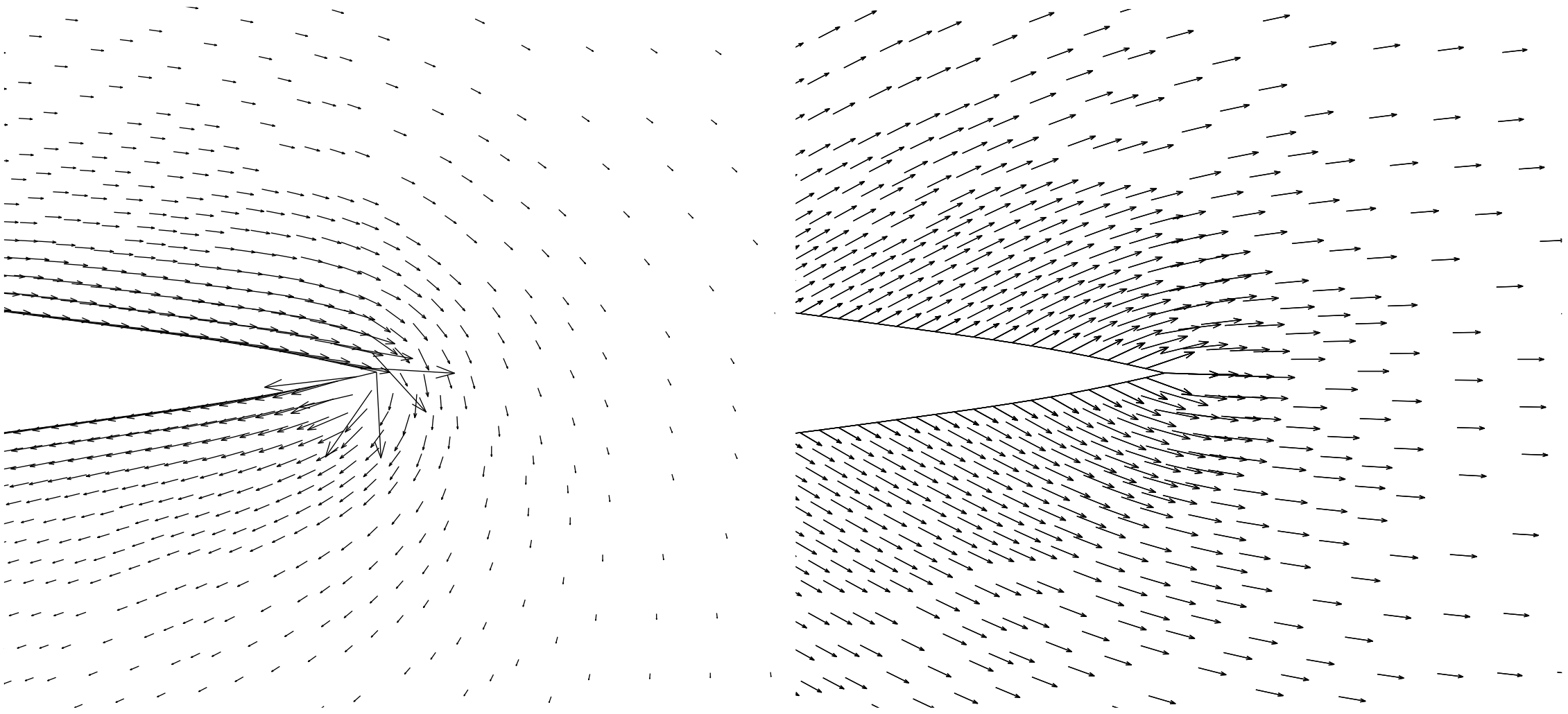

**(a)** **(b)**

**Fig. 9 Drag adjoint velocity $\vec{\varphi}=(\psi_2,\psi_3)$ near the trailing edge of a NACA0012 airfoil at flow conditions (a) $M_\infty = 0.8$, $\alpha = 1.25^o$ and (b) $M_\infty = 0.8$, $\alpha = 0^o$.**

*Subsonic flow*

In the subsonic (shockless) case, both the trailing edge singularity and the mesh divergence disappear altogether from the drag adjoint solution for any angle of attack, as can be seen in Fig. 10 and Fig. 11 .

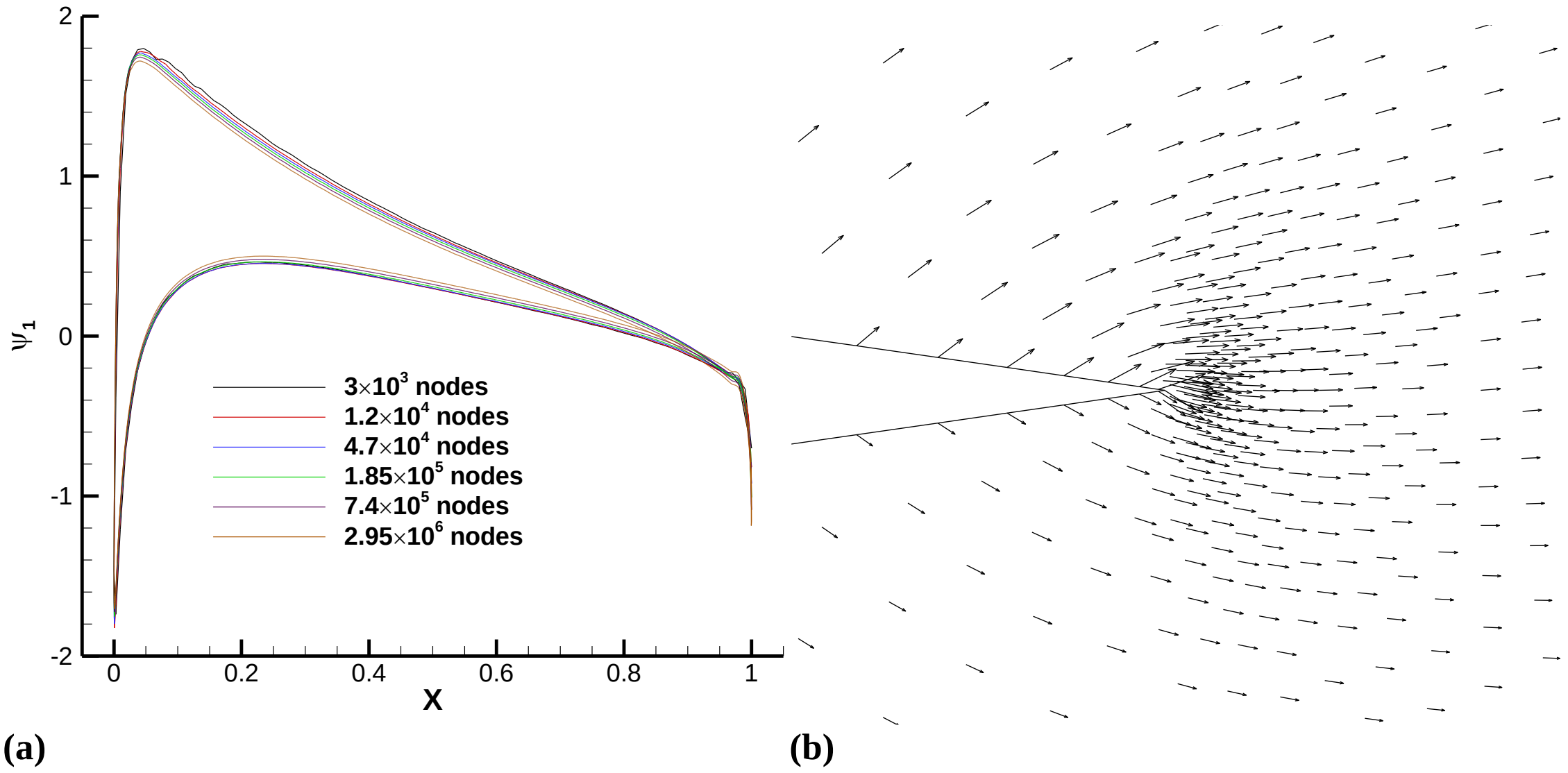

**(a)** **(b)**

**Fig. 10 Tau solver, NACA0012 with $M_\infty = 0.5$ and $\alpha = 2^o$ . (a) Drag density adjoint on the airfoil. (b) Pattern of the drag adjoint momentum vector $\vec{\varphi}=(\psi_2,\psi_3)$ near the trailing edge.**

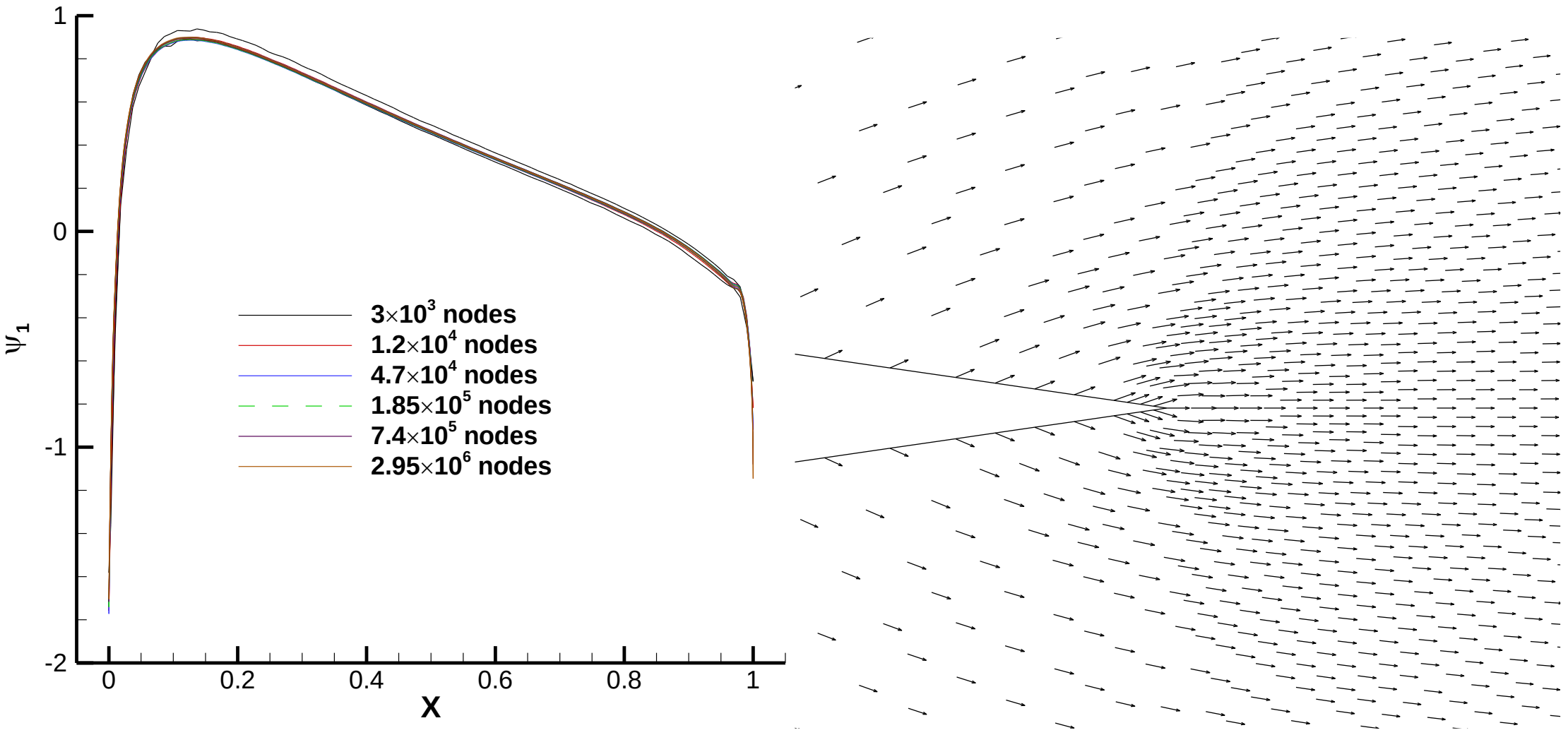

(a) (b)

**Fig. 11 Tau solver, NACA0012 with $M_\infty = 0.5$ and $\alpha = 0^o$. (a) Drag density adjoint on the airfoil. (b) Pattern of the drag adjoint momentum vector $\vec{\varphi} = (\psi_2, \psi_3)$ near the trailing edge.**

*Effect of cost function*

So far, the focus has been on aerodynamic drag. It turns out that the lift-based adjoint solution is singular at the trailing edge (and mesh divergent across the entire airfoil profile) for any flow condition, as can be seen in Fig. 12 , Fig. 13 and Fig. 14 , for flow past a NACA0012 airfoil with $M = 0.8$ , $\alpha = 0^\circ$ ; $M = 0.5$ , $\alpha = 0^\circ$ ; and $M = 0.5$ , $\alpha = 2^\circ$ , respectively.

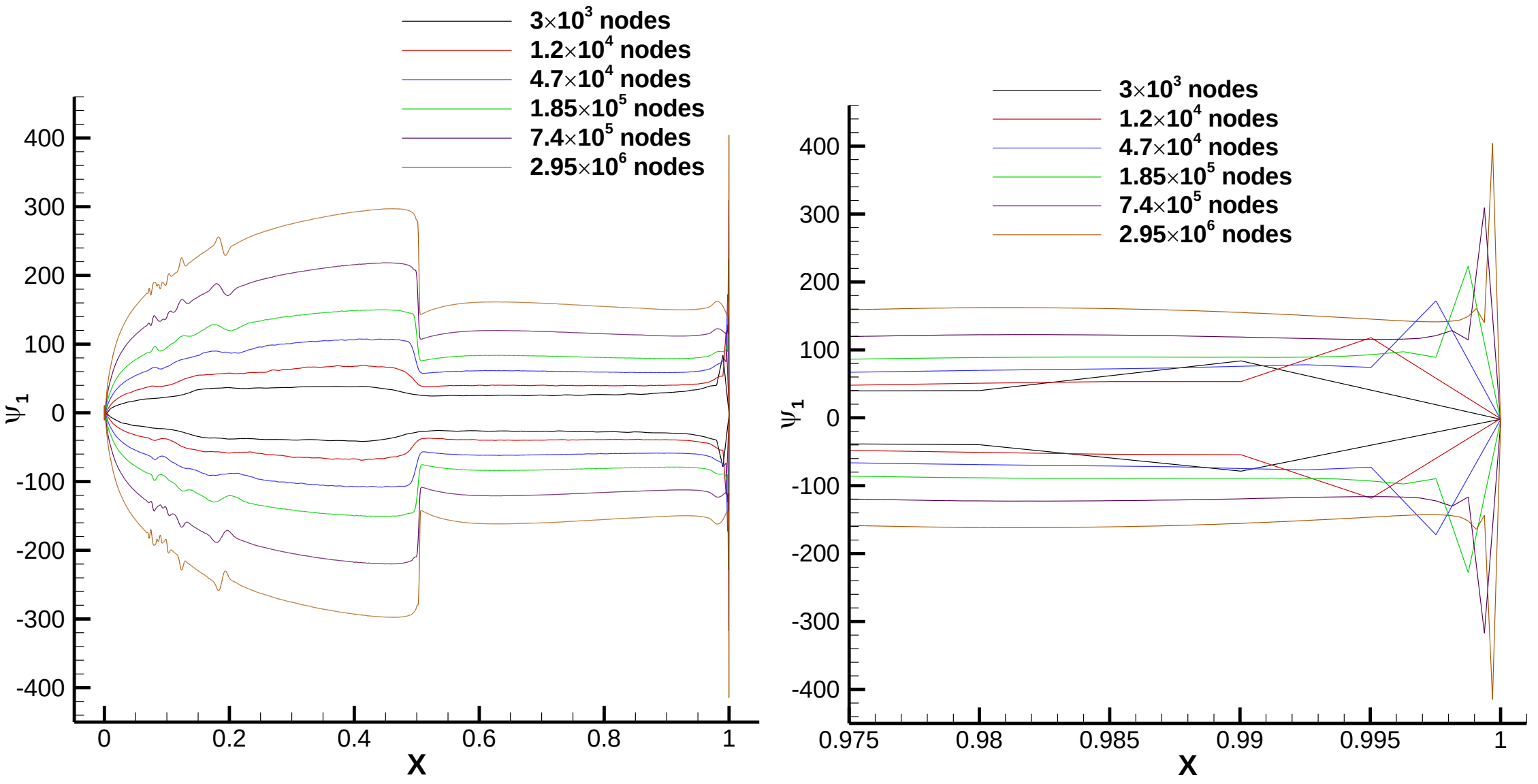

(a) (b)

**Fig. 12 Tau solver. Lift adjoint, NACA0012 airfoil with $M_\infty = 0.8$ and $\alpha = 0$. Overview (a) and zoom in the trailing edge region (b).**

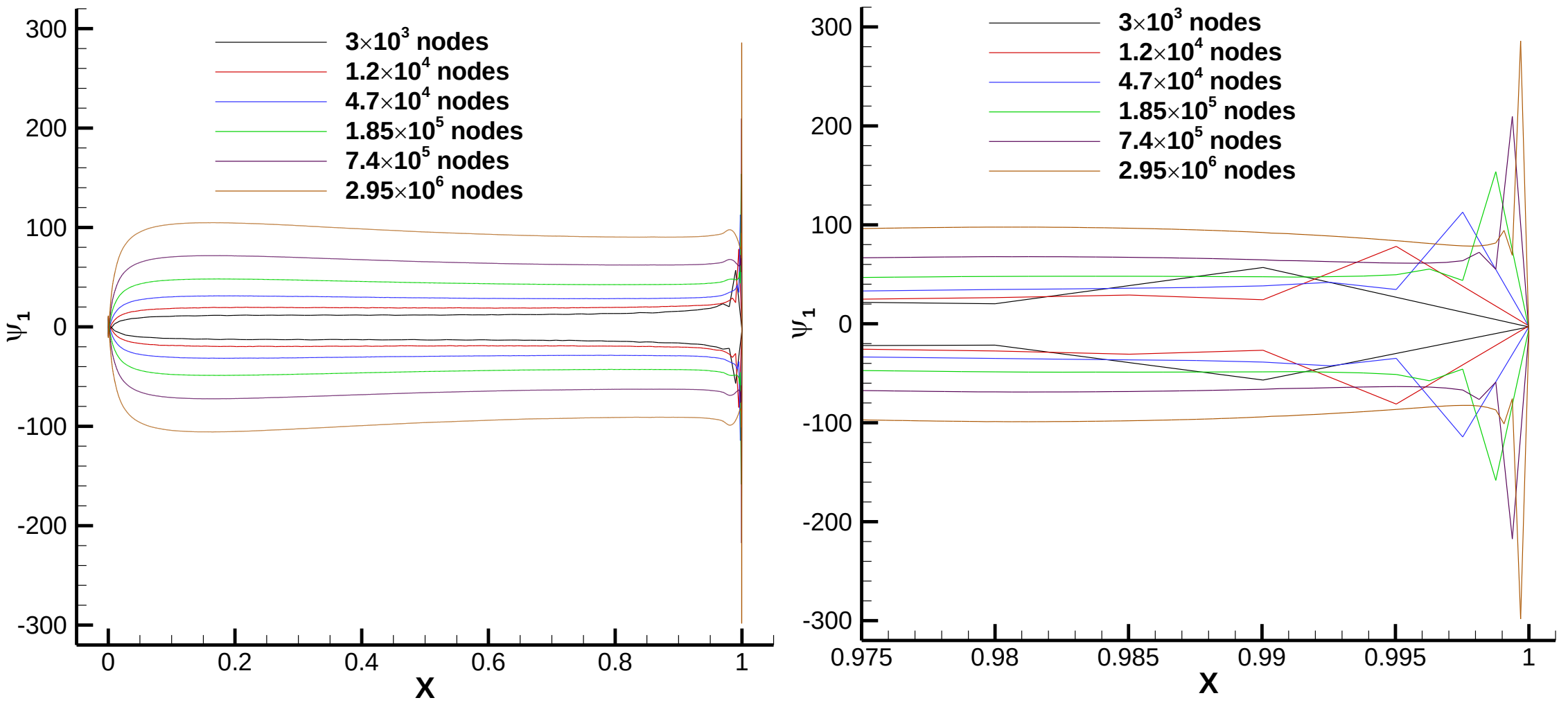

(a) (b)

**Fig. 13 Tau solver. Lift adjoint, NACA0012 airfoil with $M_\infty = 0.5$ and $\alpha = 0$. Overview (a) and zoom in the trailing edge region (b).**

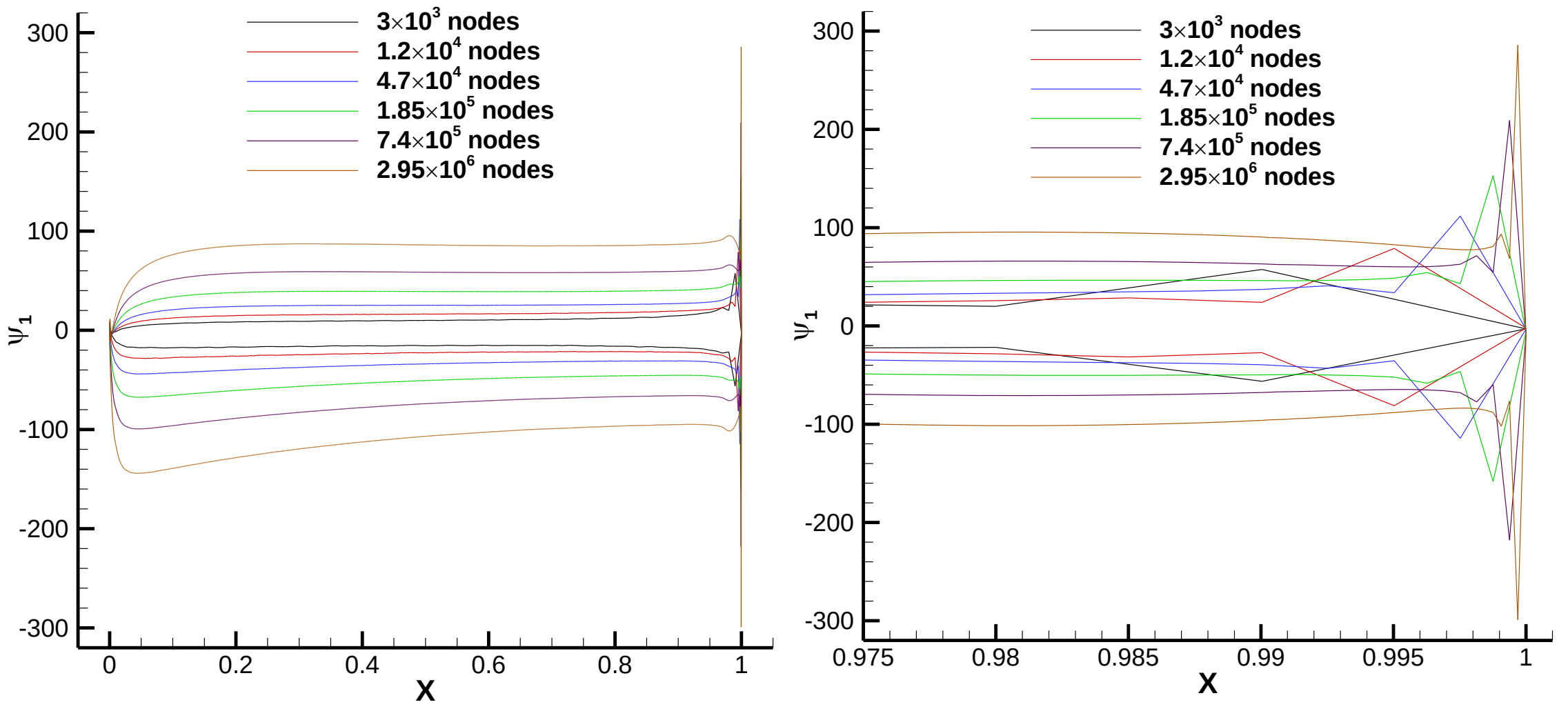

(a) (b)

**Fig. 14 NACA0012 airfoil with $M_\infty = 0.5$ and $\alpha = 2^o$. Overview of lift adjoint solution (a) and zoom in the trailing edge region (b).**

*Effect of trailing edge geometry*

Anderson and Venkatakrishnan [24] showed that the effect of the adjoint singularity at the trailing edge on the sensitivities decreases as the trailing edge angle is reduced. We may also wonder as to the behaviour of the adjoint solution when the sharp trailing edge is replaced by a finite one. To investigate these issues here a modified NACA0012 airfoil is considered at the same Mach number and angle of attack as before and with wedge angles $w = 10.68^o$ and $3.24^o$ (the initial geometry has $w \approx 20.22^o$). We also compute the symmetric Joukowski airfoil defined in section II.B and a NACA0012 airfoil with a blunt trailing edge of 0.252% chord. The results of the computations are presented in Fig. 15–Fig. 18. In all cases, the adjoint singularity is still present, along with a significant level of mesh dependence of the adjoint solutions throughout the airfoil profiles, the specific details differing for each case. In the *w* = 10.68º case (Fig. 15), the behaviour is very similar to the original NACA0012 case, with wall adjoint values growing continually as the mesh is refined. On the other hand, for *w* = 3.24º (Fig. 16) and the Joukowski airfoil (Fig. 17), the character of the solution changes and the wall adjoint values do not diverge with mesh refinement but still fail to converge even on the finest meshes throughout most of the airfoil. The only exception occurs on the suction side upstream of the primal shock, where the adjoint solution seems to converge as the mesh is refined.

Finally, for the blunt trailing edge case (Fig. 18), there is still a significant level of mesh dependence, even though only 4 mesh levels can be shown. With sufficient mesh resolution, a separation with vortex shedding and strong pressure oscillations sets in, changing the character of the solution and making it impossible to obtain converged steady adjoint solutions.

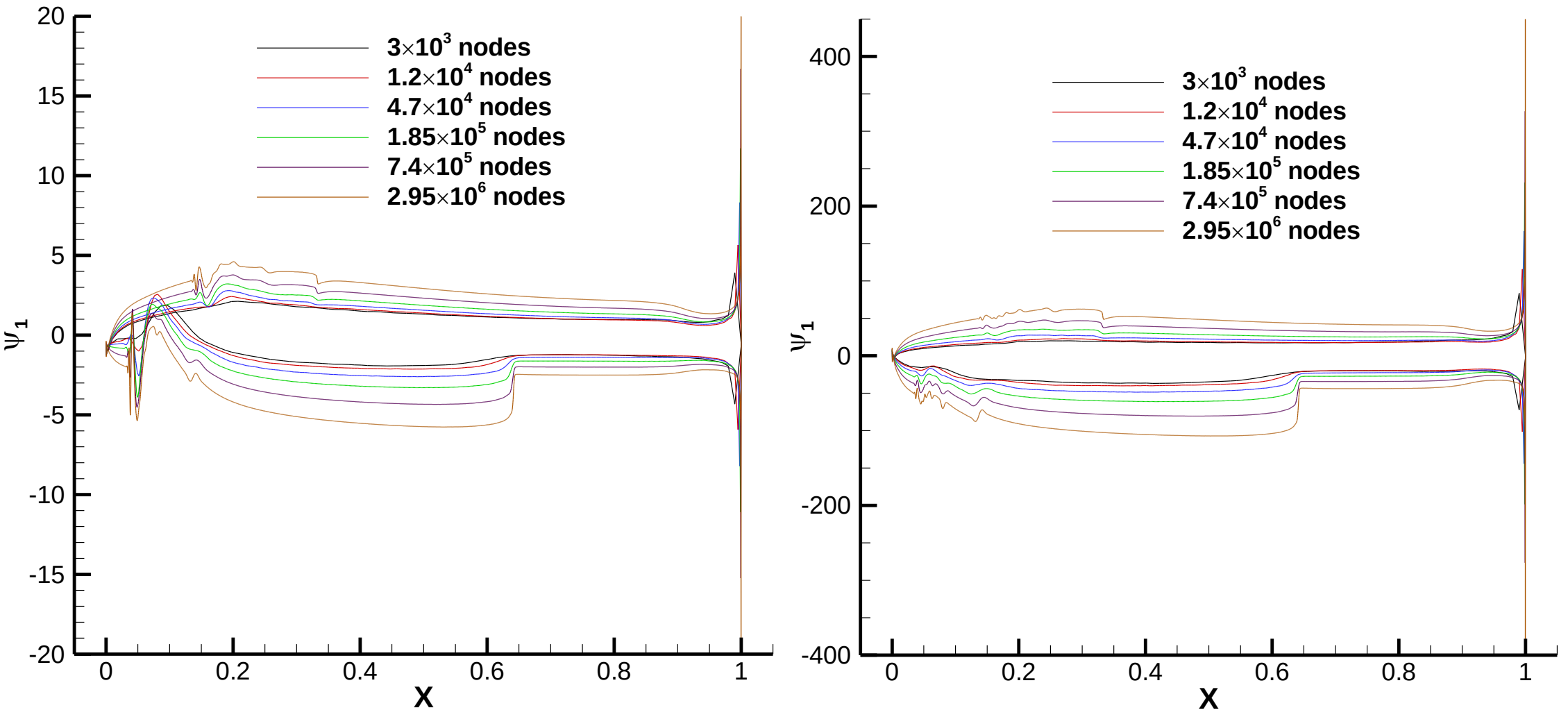

**(a)** **(b)**

**Fig. 15 NACA0012 airfoil, wedge angle 10.68º, $M_\infty = 0.8$ and $\alpha = 1.25$ (Tau solver). (a) drag adjoint solution on the airfoil profile. (b) Lift adjoint.**

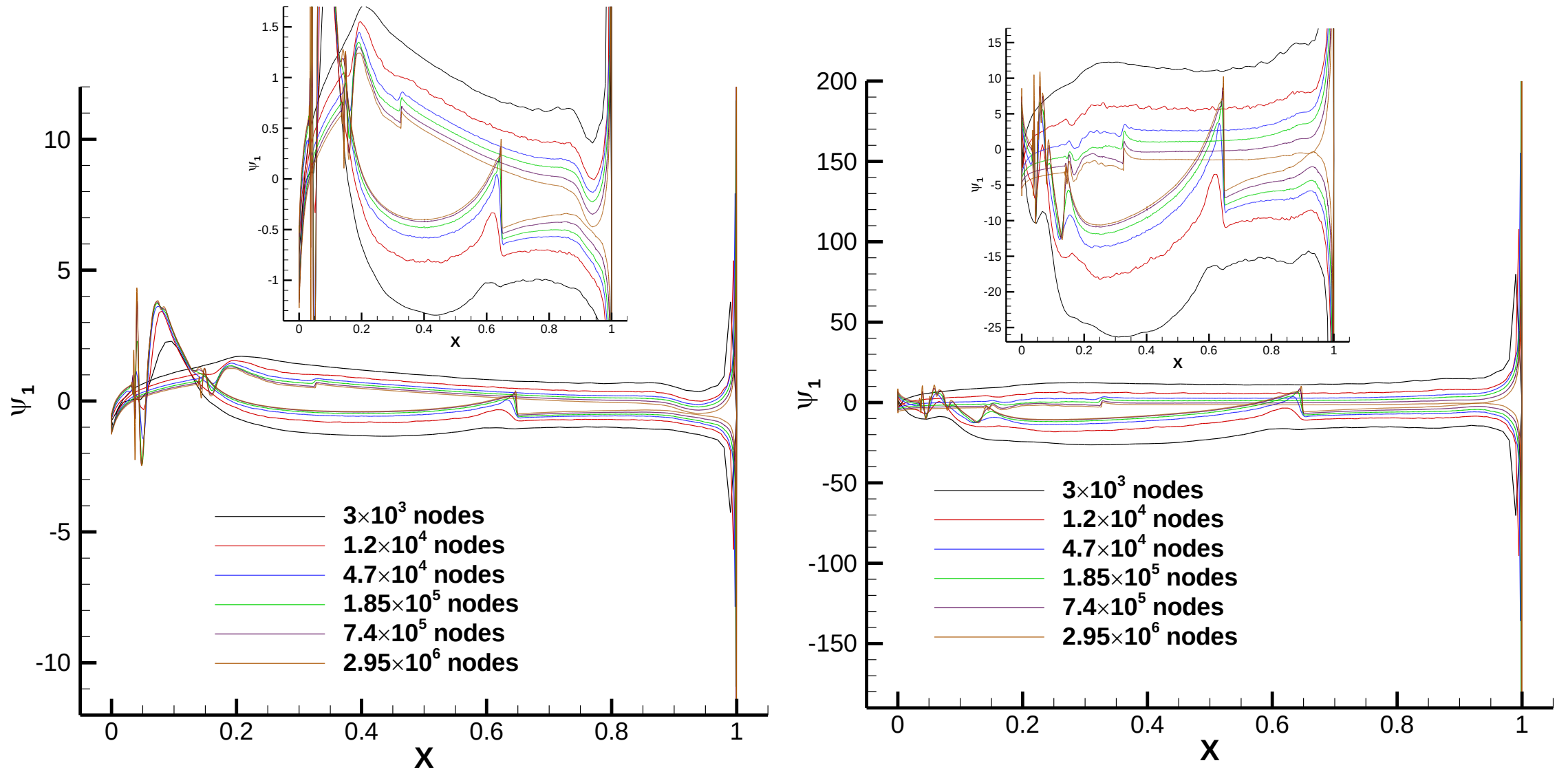

(a) (b)

**Fig. 16 NACA0012 airfoil, wedge angle 3.24º, $M_\infty = 0.8$ and $\alpha = 1.25$ (Tau solver). (a) drag adjoint solution on the airfoil profile and close –up view. (b) Lift adjoint and close-up view.**

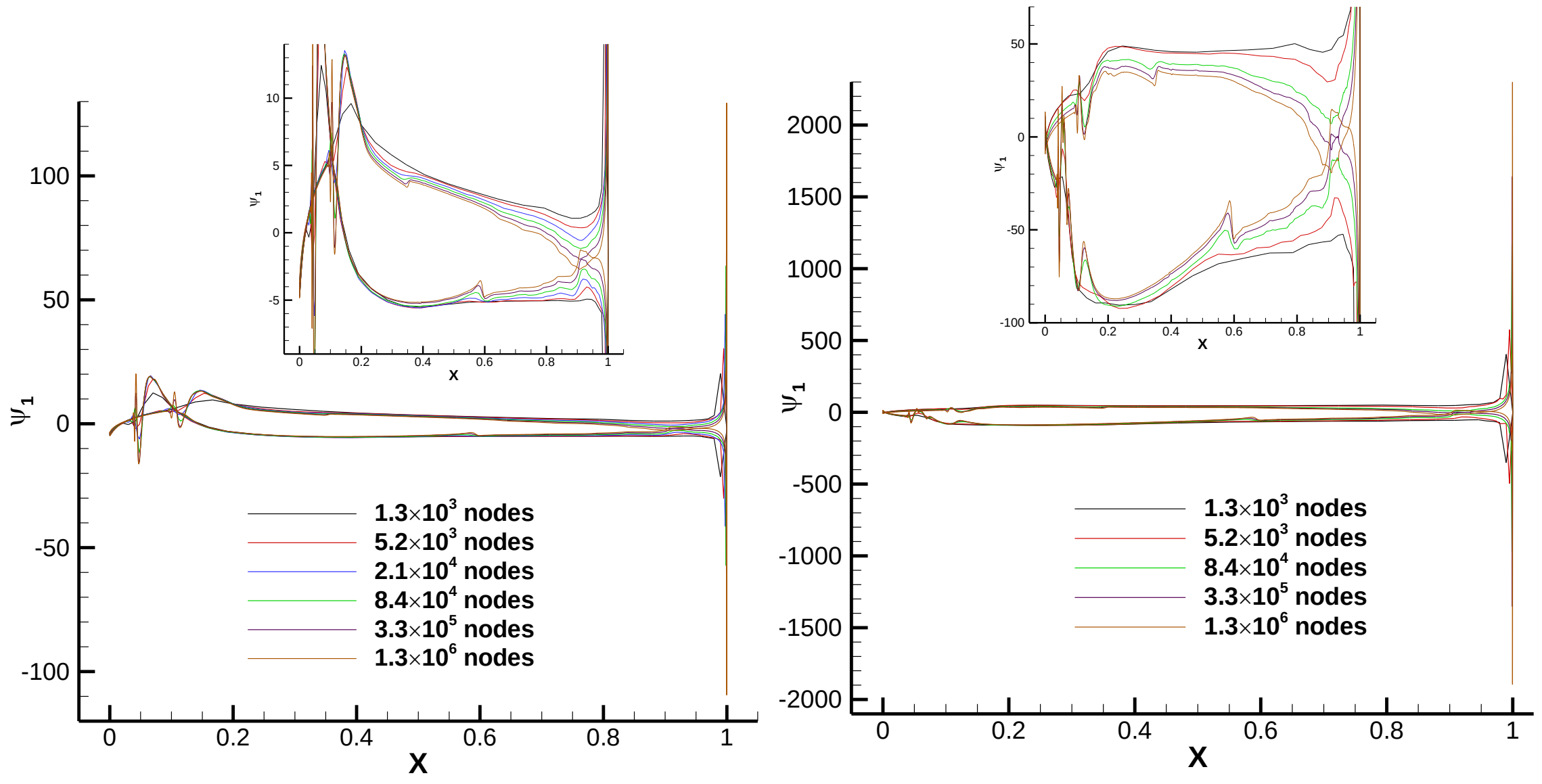

(a) (b)

**Fig. 17 Symmetric Joukowski airfoil, $M_\infty = 0.8$ and $\alpha = 1.25$ (Tau solver). (a) drag adjoint solution on the airfoil profile and close–up view. (b) Lift adjoint on the profile and close-up view.**

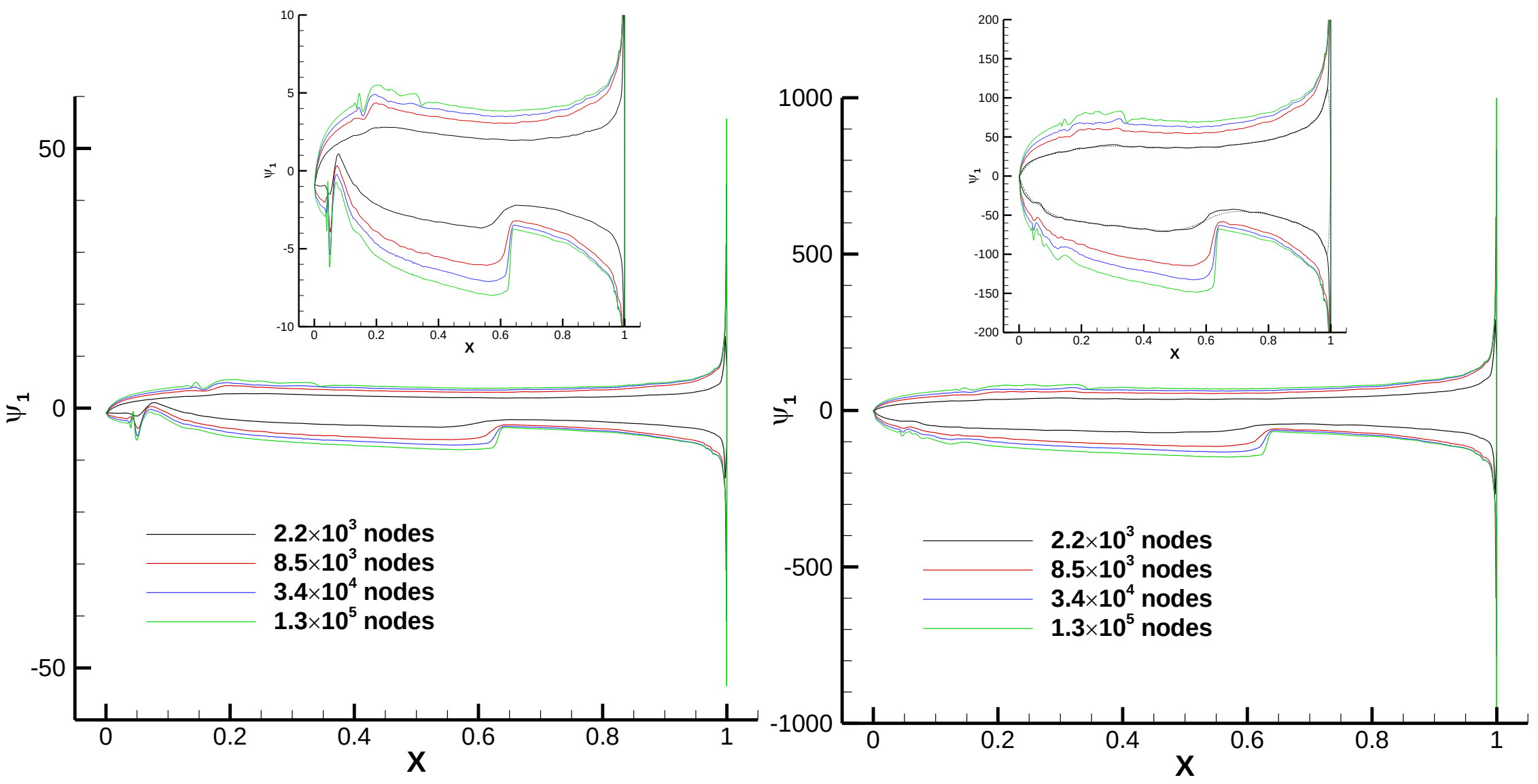

**(a)** **(b)**

**Fig. 18 NACA0012 airfoil with blunt trailing edge.** $M_\infty = 0.8$ **and** $\alpha = 1.25$ **(Tau solver). (a) drag adjoint solution on the airfoil profile and close-up view. (b) Lift adjoint and close-up view.**

*Effect of dissipation*

We can also wonder as to the effect of dissipation on the observed behavior. Following the ideas in [32], we can increase the dissipation (this can be done easily in schemes such as JST where the dissipation is scaled independently) as the mesh is refined in such a way that discontinuities such as shocks are increasingly resolved[4]. As mesh refinement proceeds by edge bisection (whereby the typical mesh spacing is halved at each stage), this is achieved by scaling the JST coefficients by $2^{k/2}$, where $k = 0,1,2,...$ and $k = 0$ corresponds to the coarsest mesh. Unfortunately, while the overall levels of the adjoint solution on the airfoil are reduced, the continuous variation with mesh density persists (Fig. 19).

[4] For the JST scheme, the effective viscosity is related to the coefficient of the second-order dissipation term $\varepsilon_2$ as $\mu_{eff} = \varepsilon_2 \ell$ where $\ell$ is, broadly speaking, the local mesh spacing. The physical width of a captured shock is proportional to $\mu_{eff}$, while the number of cells within the shock layer is proportional to $\varepsilon_2$. We need thus to scale $\varepsilon_2$ with the mesh spacing in such a way that $\varepsilon_2$ grows as the mesh is refined while $\varepsilon_2 \ell \to 0$.

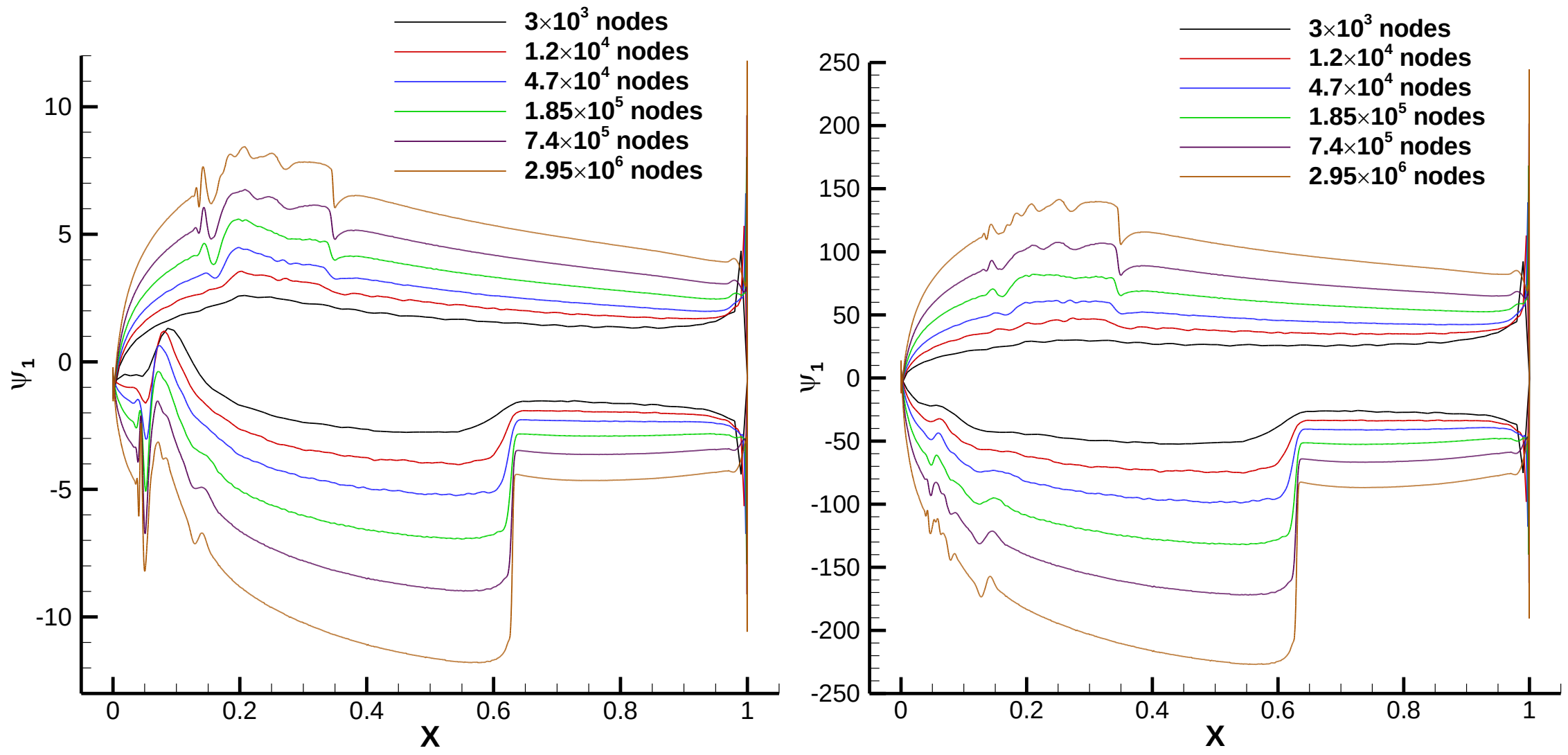

(a) (b)

**Fig. 19 NACA0012 airfoil with $M_\infty = 0.8$ and $\alpha = 1.25$ (Tau solver). JST scheme with increased dissipation (a) Drag adjoint solution. (b) Lift adjoint solution**

Another possibility would be to activate the second dissipation near adjoint discontinuities/singularities. In the JST case, this can be accomplished by adding to the pressure-based JST shock switch an adjoint-targeting term (containing, for example, the Laplacian of the adjoint density). Such adjoint-tailored dissipation does result in smoother transitions at adjoint discontinuities [17], but it is completely unnatural from the discrete adjoint viewpoint and will not be pursued here.

As an alternative, we can activate the second dissipation (for both the primal and adjoint solvers) throughout the whole computational domain. This reduces the accuracy of the solution and does not prevent the mesh dependence of the adjoint solution, as can be seen in Fig. 20, where results with everywhere active second-order dissipation, both constant and increasing with the mesh density, are shown.

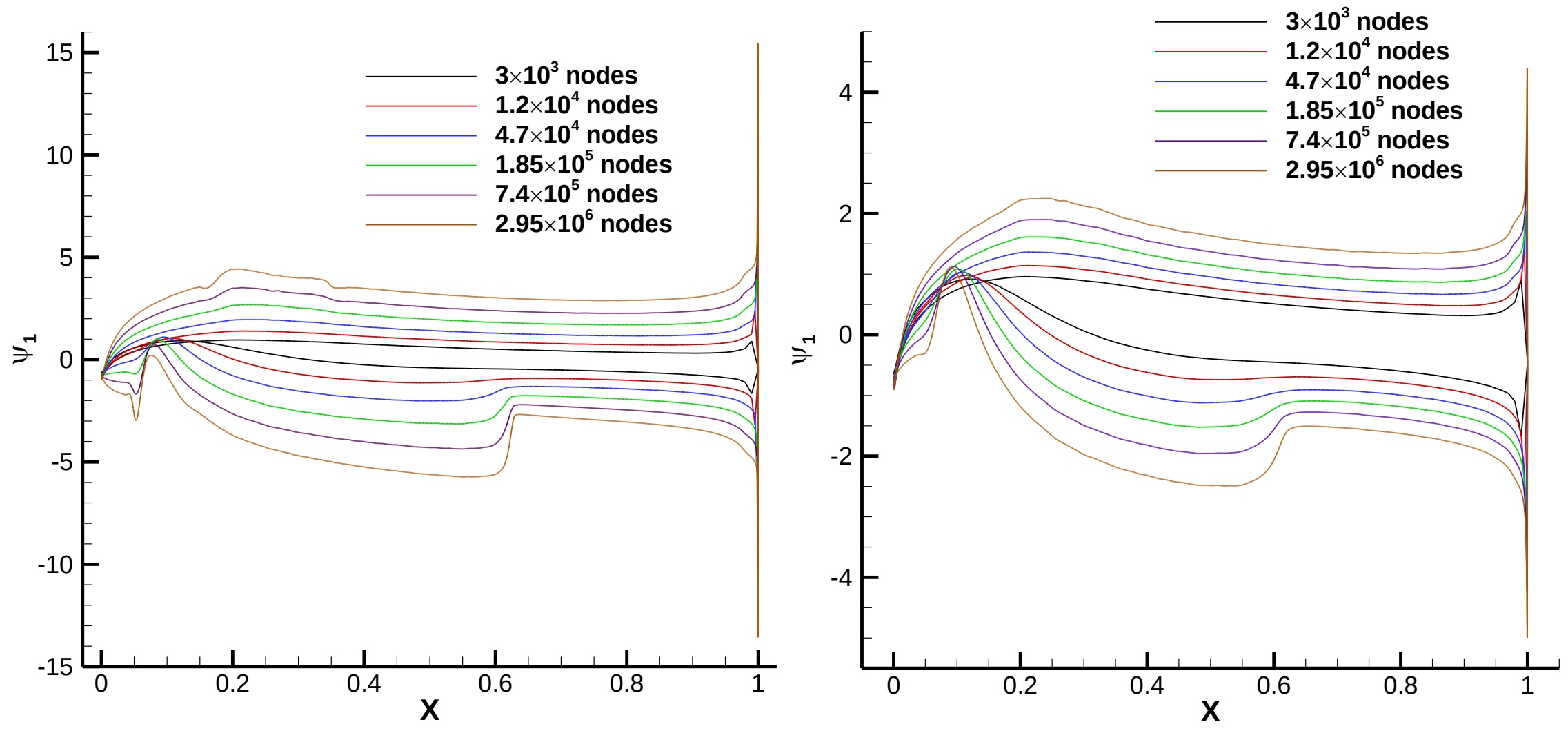

(a) (b)

**Fig. 20 NACA0012 airfoil ($M_\infty = 0.8$, $\alpha = 1.25$), Drag adjoint solution. JST scheme with constant 2nd order dissipation. (a) Fixed viscosity. (b) Viscosity increasing with mesh density.**

*Effect of computational meshes*

As mentioned above, additional computations have been carried out on a different set of isotropically refined meshes with a different node distribution on the airfoil profile. The baseline mesh has about 5200 nodes, of which 200 are on the airfoil profile and are clustered towards the leading and trailing edges to provide more resolution. The resulting computations are shown in Fig. 21. In the new set of meshes, nodes are closer to the trailing edge. Accordingly, the adjoint attains considerably higher values towards the trailing edge compared to the results in Fig. 3.

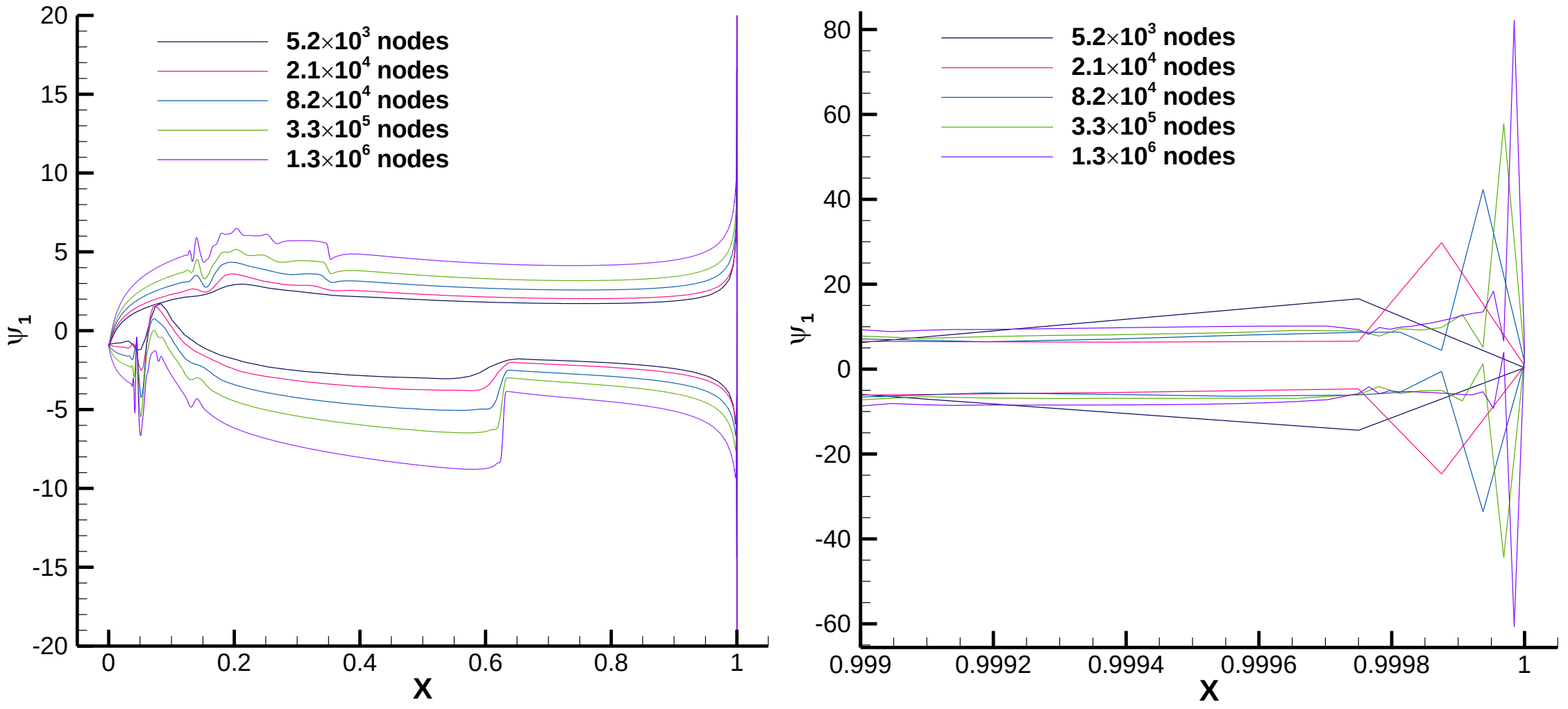

**Fig. 21 NACA0012 airfoil ($M_\infty = 0.8$, $\alpha = 1.25$). Drag adjoint density solution on the airfoil profile on a new set of meshes with the Tau solver. Overview (a) and zoom in the trailing edge region (b).**

## B. 3D inviscid flows

Next, a three-dimensional case involving transonic flow past an ONERA M6 wing [44] with $M_\infty = 0.8395$, $\alpha = 0°$ and $\alpha = 3.06°$ (and sideslip angle $\beta = 0°$ in both cases) is considered. At the second flow condition there is a "lambda" shock along the upper surface of the wing. Computations have been carried out with the discrete adjoint Tau solver on four sequentially refined tetrahedral meshes of sizes ranging from 205000 to 105 million elements, respectively.

Fig. 22 shows the drag density discrete adjoint distribution at the half-span wing section. We clearly see that, again, in the lifting case the adjoint is singular at the trailing edge, with the accompanying mesh dependence across the entire section, while both the singularity and mesh dependence are absent in the non-lifting case. In the latter case there is a mesh divergence near the leading edge that actually corresponds to the suction peak and sonic line of the primal flow as can be seen in Fig. 23.

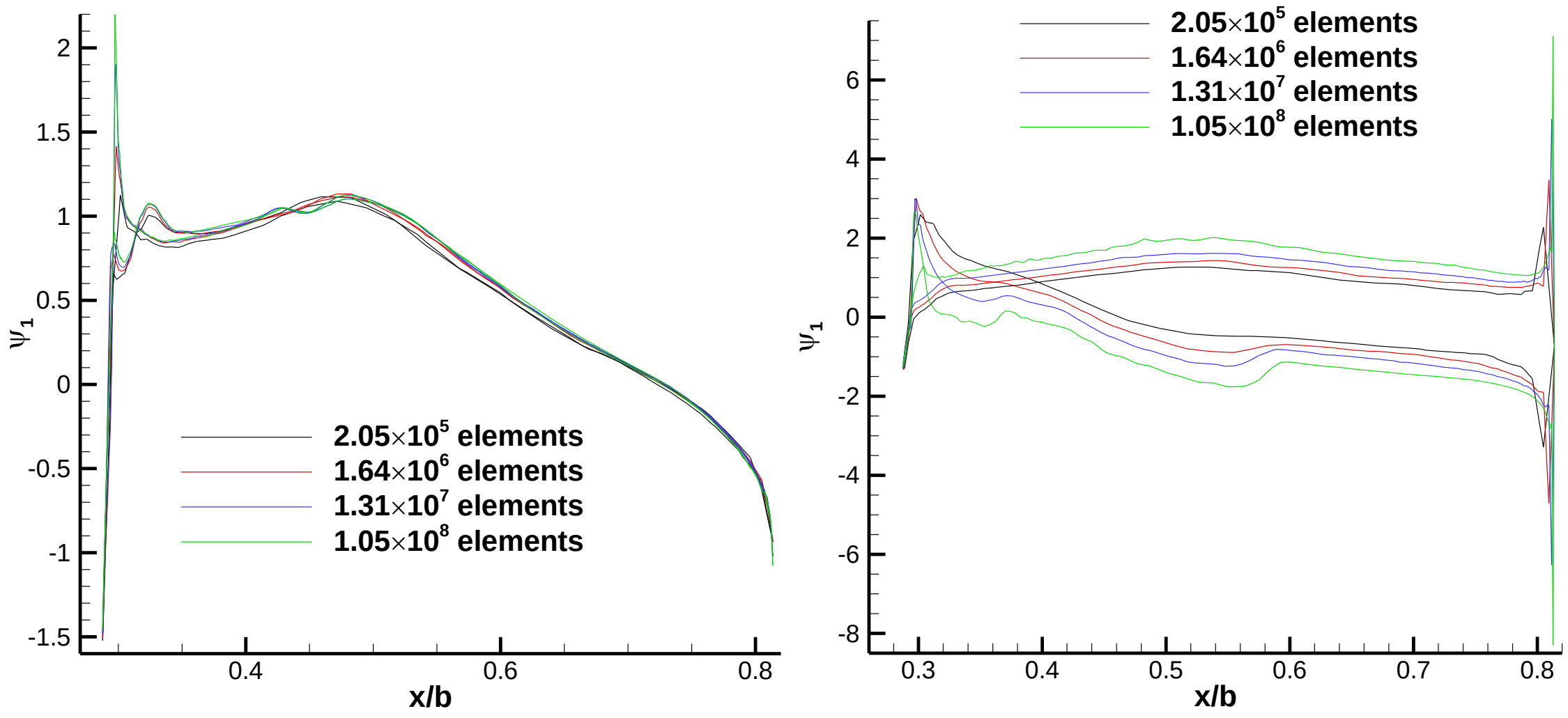

**(a)** **(b)**

**Fig. 22 Drag adjoint solution for an ONERA M6 wing with** $M_\infty = 0.8395$ **and (a)** $\alpha = 0^\mathrm{o}$ **and (b)** $\alpha = 3.06^\mathrm{o}$**.**

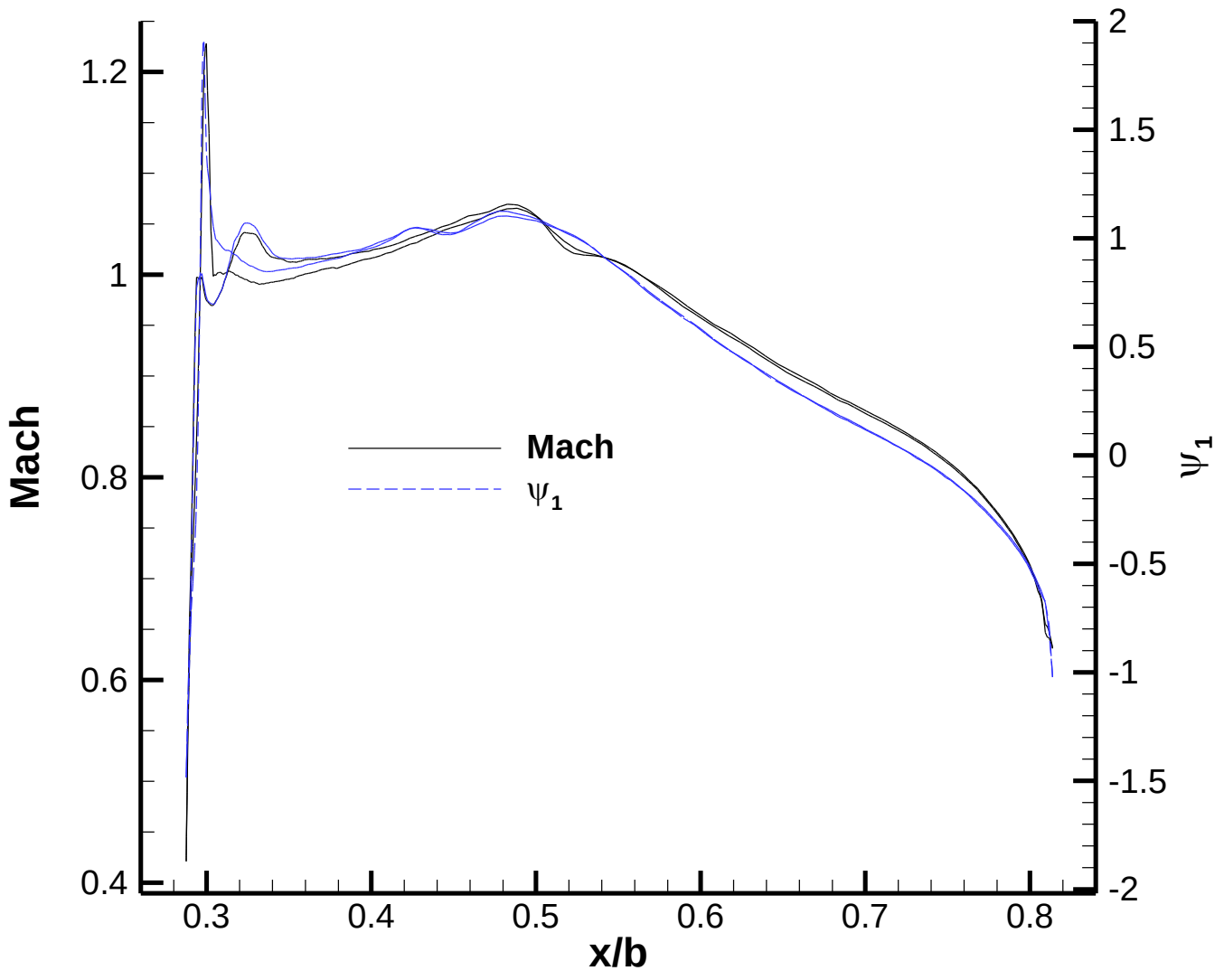

**Fig. 23 Mach number distribution vs drag adjoint solution for an ONERA M6 wing with** $M_\infty = 0.8395$ **and** $\alpha = 0^\mathrm{o}$**.**

On the other hand, the lift adjoint solution (Fig. 24) is singular at the trailing edge (and not mesh convergent across the whole section) for both flow conditions as in 2D cases.

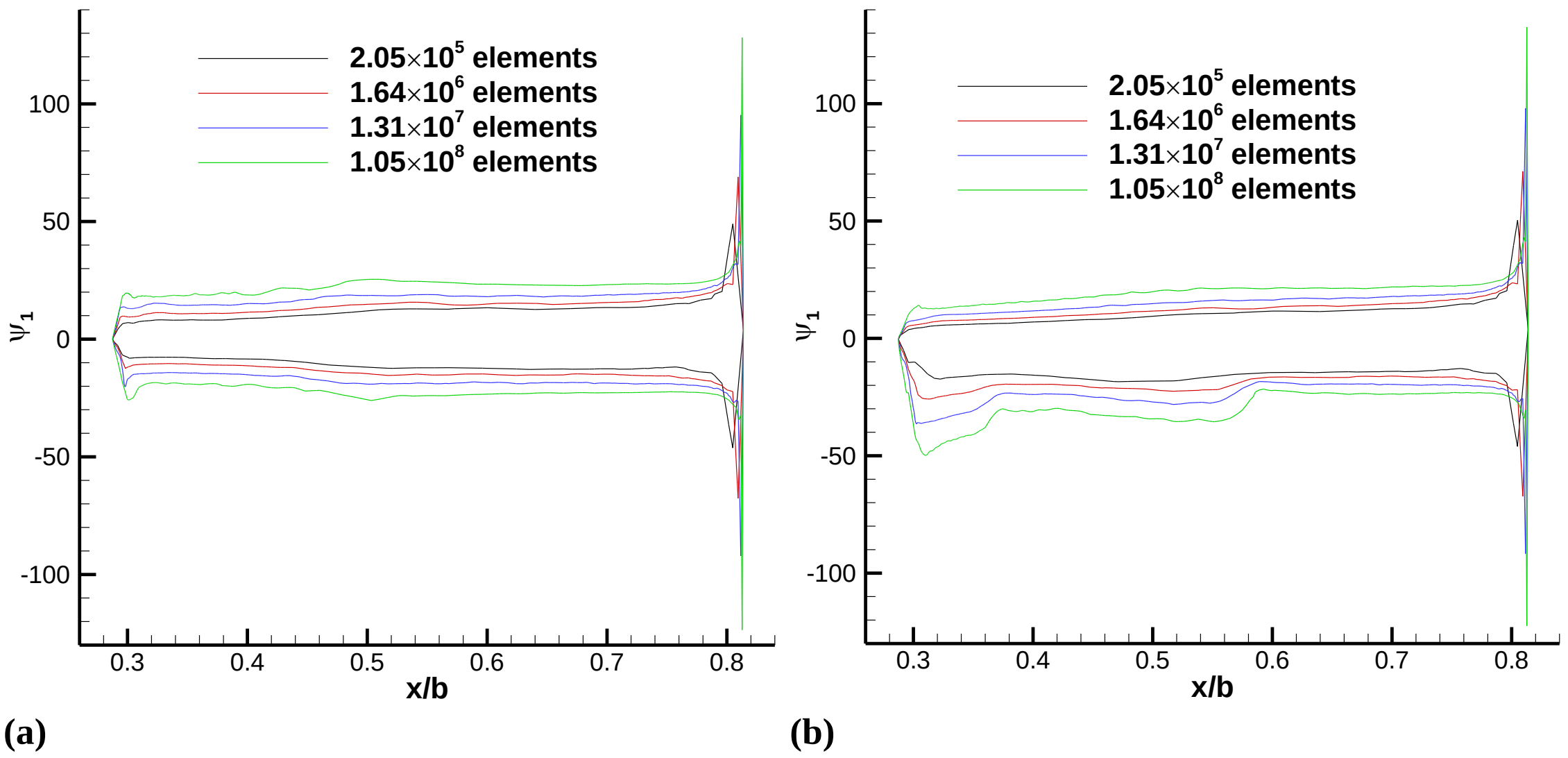

**Fig. 24 Lift adjoint solution for an ONERA M6 wing with $M_\infty = 0.8395$ and (a) $\alpha = 0^{\circ}$ and (b) $\alpha = 3.06^{\circ}$.**

## C. Viscous flows

It remains to see whether the previously described phenomenon is exclusive to the adjoint Euler equations. We shall just analyze a simple viscous case to see whether comparable issues occur. Now, with adjoint viscous solutions there is the issue that discrete adjoint solutions approaching viscous walls are typically non-smooth and oscillatory [29] (unless the discretization is dual consistent) and can even have completely arbitrary values decoupled from the interior adjoint [24]. This can obscure the problem at hand, so we will compute the numerical adjoint solution with a continuous adjoint discretization studied and validated (by means of boundary sensitivity derivatives) in [10].

We choose viscous flow past a NACA 0012 airfoil with $M_\infty = 0.5$, $\alpha = 2^{\circ}$ and Reynolds number *Re* = 5000. Computations are carried out on a set of sequentially refined hybrid unstructured meshes combining 30 structured layers of quadrilaterals in the boundary layer embedded within a triangular mesh. Wall spacing is $\Delta y_{\min} = 1.35\times10^{-4}$ in chord units, and it is kept fixed throughout the experiment, although computations on a mesh with wall spacing half of the above have also been included for completeness. The far-field boundary is approximately 100 chord-lengths away from the airfoil. Adaptation in the triangular region proceeds as explained above, while adaptation in the boundary layer proceeds by edge bisection in the streamwise (i.e., parallel to the airfoil surface) direction only.

Fig. 25 and Fig. 26 show the results of the computations. Even though there is some degree of mesh dependence, the solutions seem to slowly approach a grid converged solution.

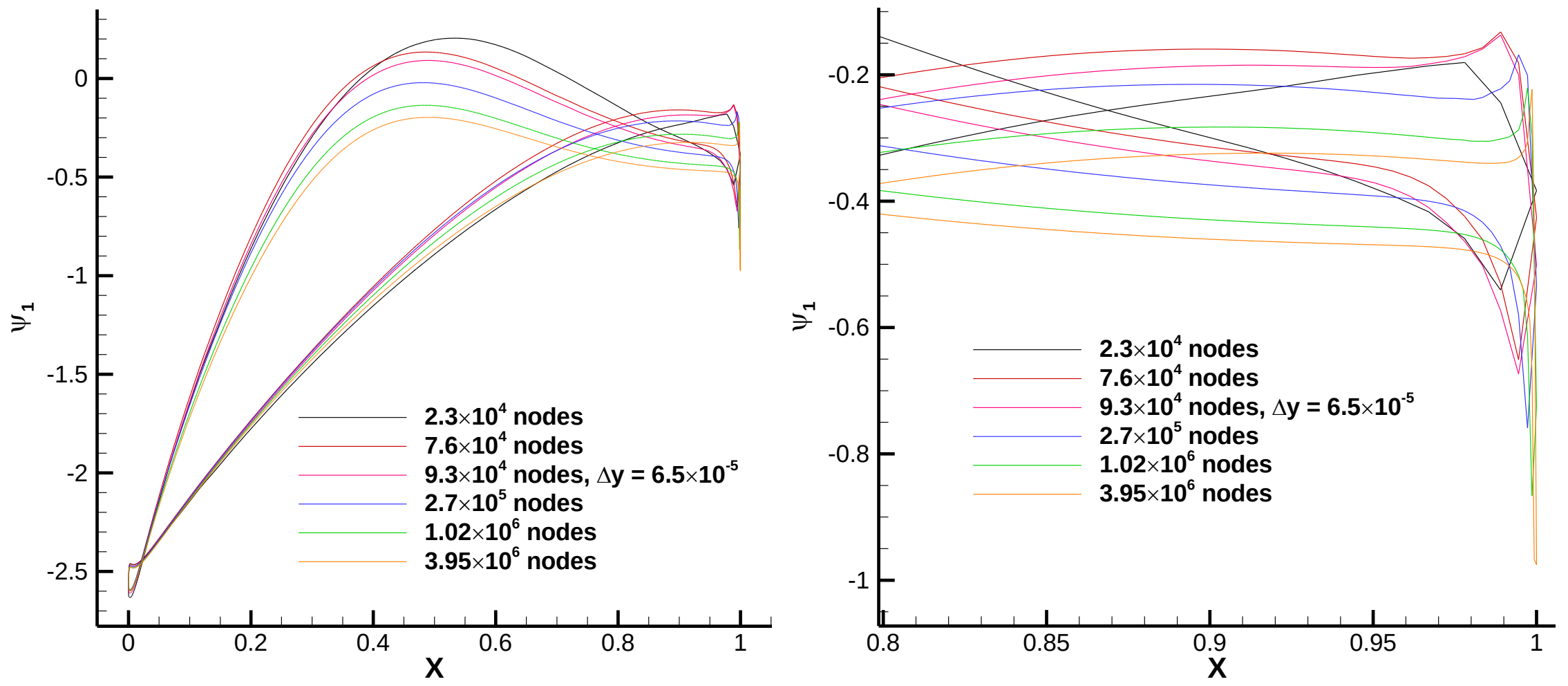

**(a)** **(b)**

**Fig. 25 NACA0012 airfoil with *Re* = 5000,** $M_\infty = 0.8$ **and** $\alpha = 2^o$. **Continuous drag adjoint solution on the airfoil surface. (a) Overview and (b) zoom around the trailing edge.**

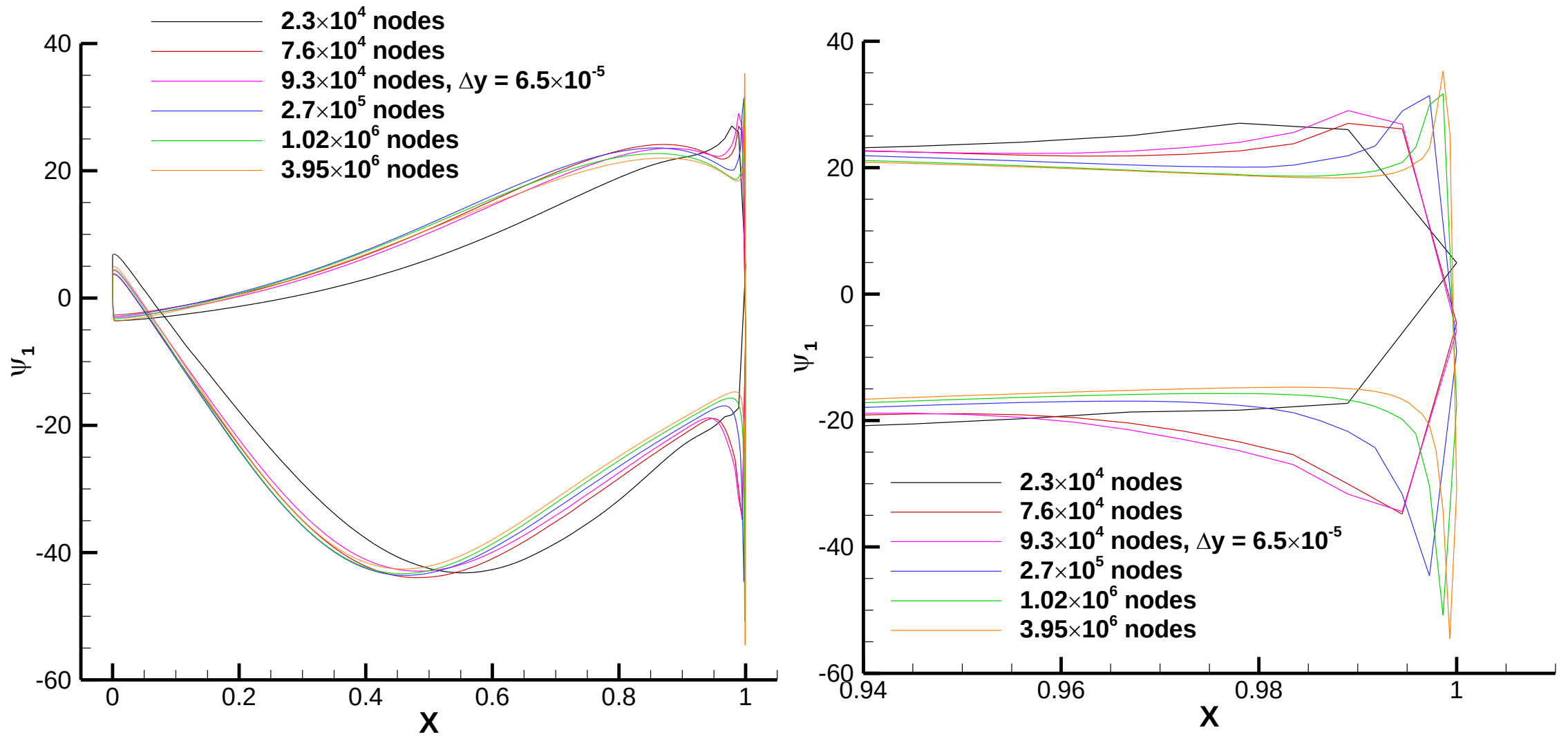

**(a)** **(b)**

**Fig. 26 NACA0012 airfoil with *Re* = 5000,** $M_\infty = 0.8$ **and** $\alpha = 2^o$. **Continuous lift adjoint solution on the airfoil surface. (a) Overview and (b) zoom around the trailing edge.**

In turbulent cases, again, we would need a fully turbulent continuous adjoint solver to investigate the behavior of the wall adjoint solution with mesh refinement. Unfortunately, neither Tau nor SU2 possess such capability, so we can only speculate that, as in the laminar case, the flow/adjoint wall b.c. is crucial here to prevent the anomalous behavior.

### D. Discussion

Section III has been devoted to characterizing the issue of mesh convergence of adjoint solutions. The problem is real and appears to affect inviscid solutions only. While this is of no immediate practical relevance, as adjoint applications focus now on viscous applications, it is still important, from a fundamental viewpoint, to fully understand the behavior of inviscid adjoint solutions for both validation of numerical solvers and a deeper understanding of the adjoint equations. We thus need to find the source of the problem, which has been found to be clearly related to the adjoint singularity at the trailing edge and to affect different cost functions in different ways. Lift adjoint solutions are affected for any flow condition, while drag adjoint solutions seem to be affected only in those cases where the flow conditions result in the formation of a slip line emanating from the trailing edge. There are several possible explanations for this behavior one can come up with.

(1) The first possibility is that the adjoint p.d.e. is ill defined at solid walls, or that the adjoint solution is singular. It has been already argued that this does not seem to be the case. Hence, if we assume that the adjoint p.d.e has a well-defined analytic solution that is smooth (and has thus a finite value) at generic points of wall boundaries, the origin of the observed behavior must be numerical.

(2) The problem may arise as a result of a dual inconsistency of the numerical scheme. I believe that this possibility is ruled out by the following arguments.

- The problem depends on the cost function and on the flow conditions.

- It arises with both continuous and discrete adjoint schemes.

- It is observed in different solvers and with different numerical schemes.

(3) The observed behavior is reminiscent of what is observed with adjoint solutions in presence of primal shocks, particularly for non-linear objective functions. At primal shocks, the adjoint equation requires an internal boundary condition which is not enforced numerically and that may lead to wrong adjoint solutions unless sufficient dissipation is applied across the shock [32]. Failing to do so leads to incorrect adjoint values in the shock region that also depend strongly on the mesh density (see chapter 3 in [45] for examples). However, the problem we are dealing with also afflicts adjoint solutions for non-shocked flows, and there are shocked (non-lifting) solutions that do not show the problem. Likewise, adding more dissipation does not cure the mesh dependence, so even though the presence of the shock may contribute to the problem, it is certainly not the sole reason. In this regard, notice that in all transonic cases, Fig. 3, Fig. 12, Fig. 15, Fig. 16, Fig. 17, Fig. 18, Fig. 22, and Fig. 24, the adjoint solutions seem to develop a quasi-shock structure which becomes more prominent as the mesh density grows. We do not know the origin of this structure, as lift/drag adjoint solutions should be continuous across primal shocks (at least for cost functions that are linear functions of the pressure) [12, 13, 15, 17], but is likely to be the result of an adjoint singularity. While adjoint solutions cannot sustain shocks, they can display jump discontinuities (see [17]). Although we do not have proof of this, its origin is likely in the adjoint singularity along the supersonic characteristic emanating from the shock root described earlier.

(4) The drag adjoint solution has been found to lack mesh convergence properties (and even be mesh-divergent in several cases) for transonic lifting flows. In such cases, the Kutta condition gives rise to the creation of a slip line/surface [43], a wake contact discontinuity that is a weak solution to the Euler equations [46] and across which there is zero normal velocity and a jump in tangential velocity and total pressure. It has been also reported [36] that adjoint solutions are discontinuous across such contact discontinuities. While this type of singularity cannot be the general cause of the observed behavior, as the lift cost function is singular for any flow condition, it might shed some light on the problem for the drag cost function. However, as shown in Fig. 27, there is no singular behavior of the adjoint solutions at the slip line (not even for the drag adjoint in transonic rotational flows). Furthermore, characteristic curves are parallel to contact discontinuities and slip lines rather than impinging on them, so no adjoint b.c. needs to be imposed, thus closing the door to a possible non-uniqueness mechanism.

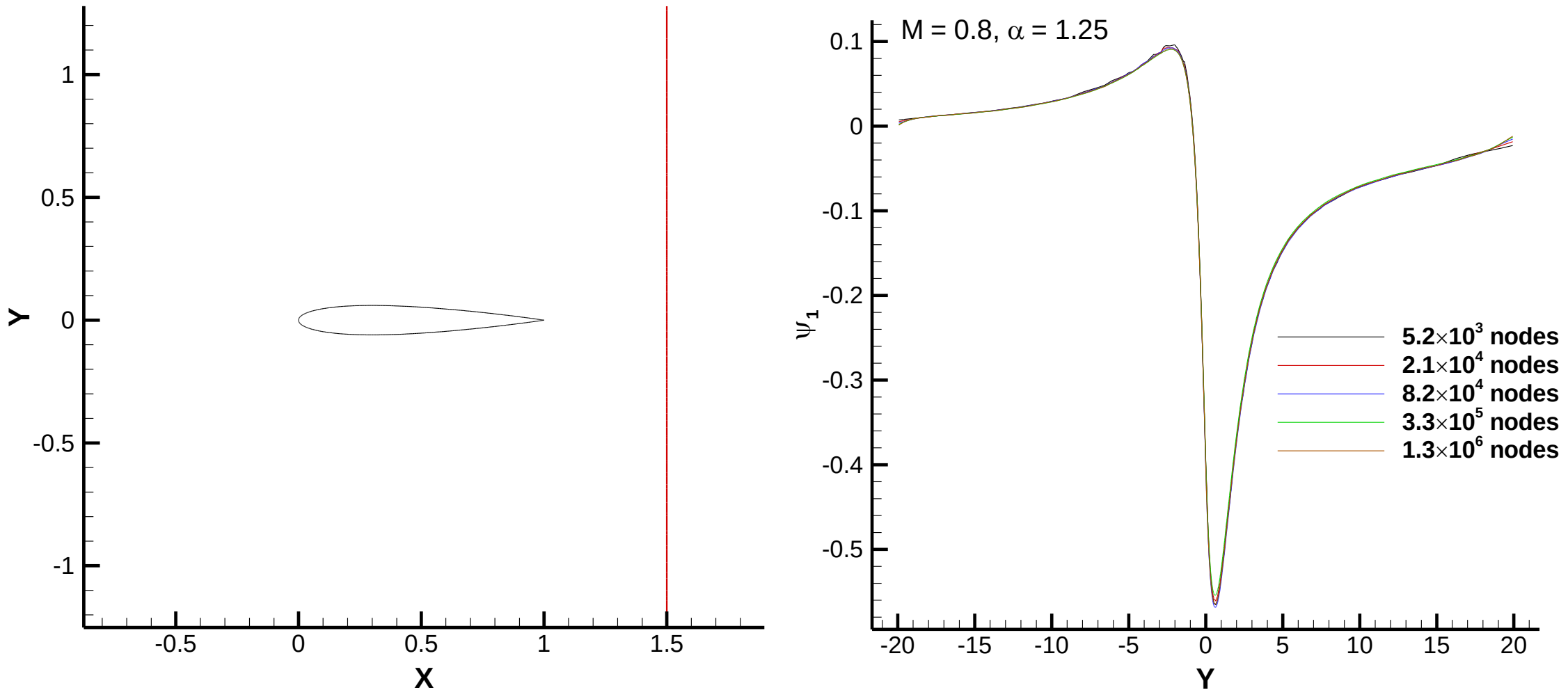

**(a)** **(b).**

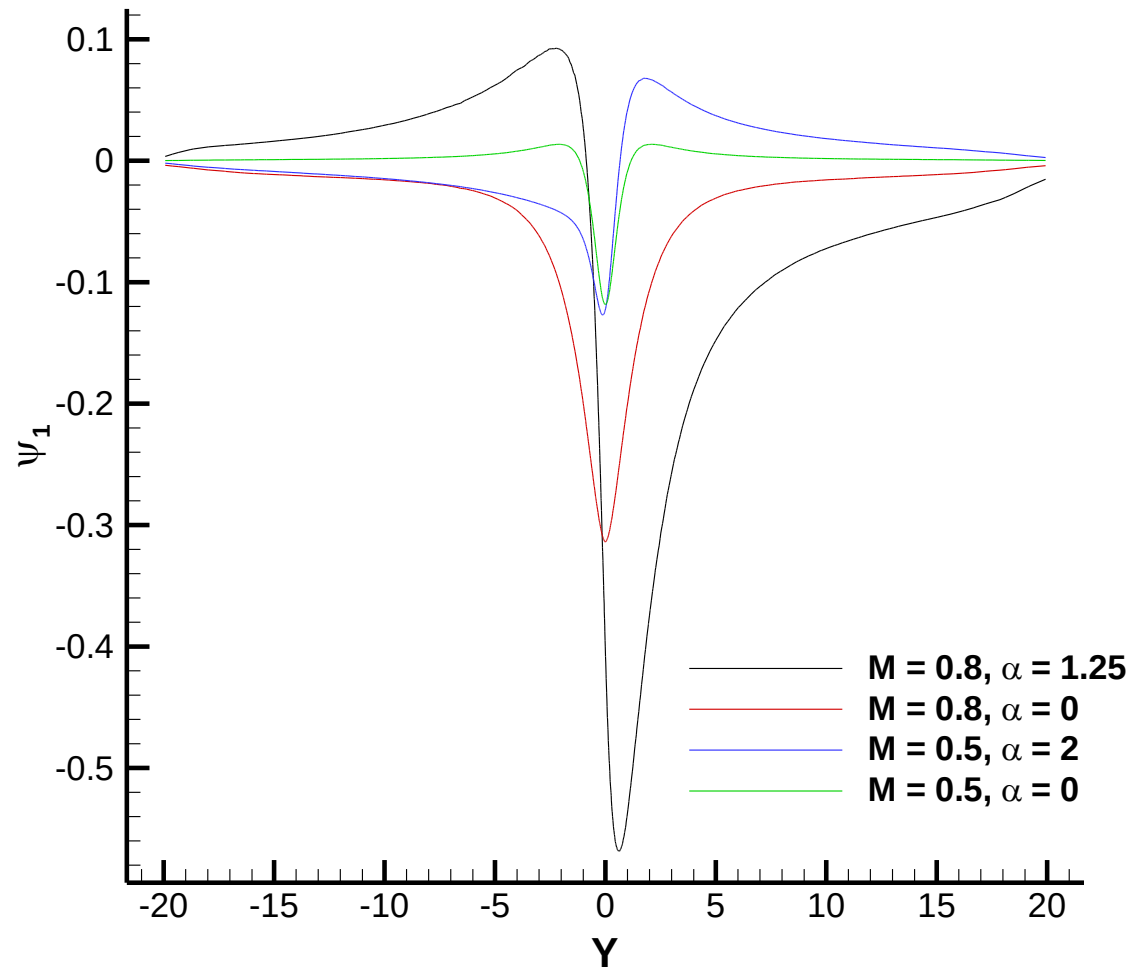

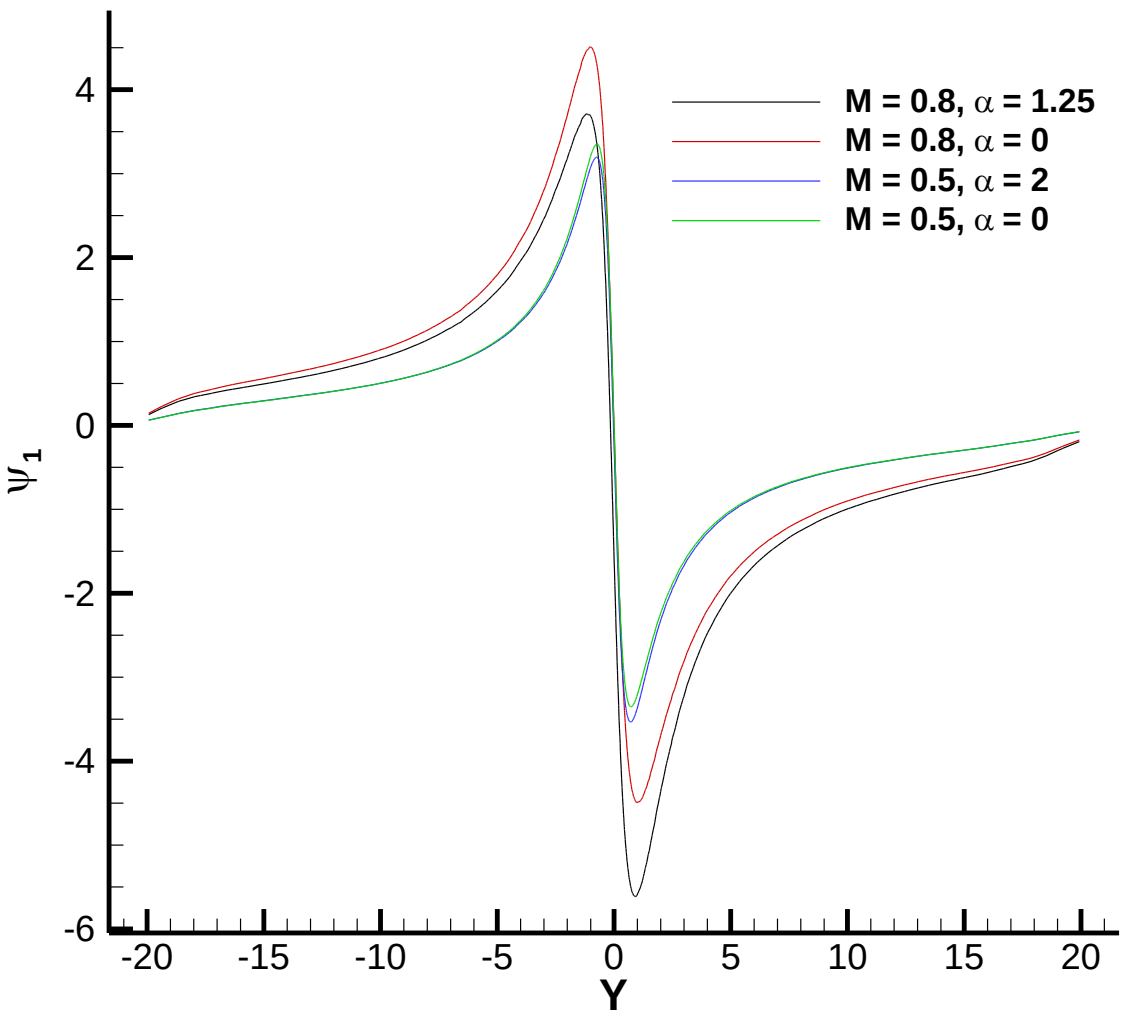

**(c)** **(d)**

**Fig. 27 NACA0012 airfoil (inviscid). Drag (b) and (c) and lift (d) density adjoint along a line crossing the wake downstream of the airfoil as indicated in (a)**

(5) Numerical effect triggered by the trailing edge singularity. What all the analyzed cases have in common is the trailing edge adjoint singularity. When the trailing edge is not singular, the adjoint values converge with mesh refinement except at certain singular points. The two phenomena are thus clearly correlated. The issue is whether the established correlation actually implies causation. Adjoint solutions with singularities are well-known, such as in quasi-1D flows, which however do not lead to a comparable level of mesh dependence of the solutions. Two and three-dimensional cases are significantly more complex, and in absence of another plausible mechanism, we are led to believe that the trailing edge singularity (which is already present in the analytic solution) does cause the mesh dependence of adjoint values at the wall. Mesh refinement adds nodes closer and closer to the trailing edge resulting in larger and larger values of the adjoint state. In turn, these large values contaminate the adjoint state across the profile and near wall regions, as only the normal value of the adjoint velocity $\vec{\varphi} = (\psi_2, \psi_3)$ is fixed by the adjoint wall boundary condition (3), triggering the observed behavior.

## IV.Conclusions

The purpose of this investigation has been to address a numerical problem observed in inviscid adjoint solutions for flows past sharp trailing edges. 2D and 3D inviscid adjoint solutions are generically singular at sharp trailing edges. This is a well-known fact that has known consequences (in both the computation of adjoint gradients and in adjoint-based error estimation and mesh adaptation) but also unexpected ones. In this paper we have reported on a new one: the lack of mesh convergence of wall and near-wall adjoint values across the entire wall boundary. We have seen that this behavior is rather generic. It is a problem of the numerical solution whose details depend on the cost function and the flow conditions, with lift adjoint variables being affected at any flow condition, while drag adjoint variables are only affected in transonic lifting cases.

This problem is not inconsequential: it makes it difficult to interpret numerical results and it can pose a problem in mesh adaptation, as the growing size of wall adjoint variables may result in excessive refinement towards the wall. On the other hand, sensitivity derivatives computed with the singular adjoint solutions are actually quite accurate and do not reflect a comparable level of mesh dependence. This observation may explain why this issue had been largely unnoticed.

What is perhaps more relevant is the fact that the lack of mesh convergence (and mesh divergence in some cases) is always associated to a singular behavior at the trailing edge. In those cases where the adjoint is not singular at the trailing edge, the mesh convergence problem is also absent, so both issues are clearly related. Since the singularity at the trailing edge is inherent in the analytic adjoint solution, we conjecture that the mesh convergence problem is actually caused by the trailing edge singularity.

All numerical computations have been performed with cell-vertex schemes. We may wonder whether switching to a cell-centered code would produce a different behavior. However, Fig. 4 in [12], which was computed with a cell-centered adjoint solver, does show the same singular behavior towards the trailing edge. While no mesh-convergence analysis is performed there, we can only speculate that the same behavior would be observed in that case as well[5].

## Funding Sources

The research described in this paper has been supported by INTA and the Ministry of Defence under the grant Termofluidodinámica (IGB99001).

## Acknowledgements

The author would like to thank J. Ponsin and D. Ekelschot for discussions. The computations reported in the paper were carried out with the TAU Code, developed at DLR's Institute of Aerodynamics and Flow

[5] One of the anonymous reviewers has confirmed having obtained the same behavior with a cell-centered, discrete adjoint code.

Technology at Göttingen and Braunschweig, which is licensed to INTA through a research and development cooperation agreement. Cross-checking has been also performed with the SU2 code, an open source code developed at the Aerospace Department of Stanford University.